\documentclass[12pt]{article}

\usepackage[fleqn]{amsmath}
\usepackage{amsfonts}
\usepackage{amssymb}
\usepackage{epsfig}

\DeclareMathOperator{\Beta}{B}

\newcommand{\drm}{\mathrm{d}}

\newcommand{\Ddt}{\frac{\drm\phantom{s}}{\drm t}}

\newcommand{\pddt}{\frac{\partial\phantom{t}}{\partial t}}
\newcommand{\pddct}{{\textstyle{\frac{1}{c}\frac{\partial\phantom{t}}{\partial t}}}}
\newcommand{\pddctSQUARE}{{\textstyle{\frac{1}{c^2}\frac{\partial^2\phantom{t}}{\partial t^2}}}}

\newcommand{\refeq}[1]{(\ref{#1})}

\newcommand{\me}{m_{\text{e}}}

\newcommand{\mbare}{m_{\text{b}}}

\newcommand{\vect}[1] {\boldsymbol{{ #1}} }

\newcommand{\qv}[1]{{\textbf{\textsl{#1}}}}

\newcommand{\tenseur}[1]{{\textbf{\textsf{#1}}}}

\newcommand{\tens}{\otimes}     

\newcommand{\Cset}{\mathbb{C}}

\newcommand{\Rset}{\mathbb{R}}

\newcommand{\II}{\tenseur{I}}           

\newcommand{\gQ}{\tenseur{g}}           

\newcommand{\uQ}{\qv{u}}                

\newcommand{\fV}{\vect{f}}              
\newcommand{\fVN}{\vect{F}}              
\newcommand{\jV}{{\vect{j}}}		
\newcommand{\nV}{{\vect{n}}}		
\newcommand{\qV}{{\vect{q}}}            
\newcommand{\QV}{{\vect{Q}}}            
\newcommand{\sV}{{\vect{s}}}            
\newcommand{\vV}{{\vect{v}}}            

\newcommand{\BLWr}{{\BV}_{\text{\textsc{lw}}}^{\text{\tiny{ret}}}} 
\newcommand{\ELWr}{{\EV}_{\text{\textsc{lw}}}^{\text{\tiny{ret}}}} 
\newcommand{\BLWa}{{\BV}_{\text{\textsc{lw}}}^{\text{\tiny{adv}}}} 
\newcommand{\ELWa}{{\EV}_{\text{\textsc{lw}}}^{\text{\tiny{adv}}}} 

\newcommand{\BVM}{{\BV}_{\text{\textsc{m}}}}    
\newcommand{\EVM}{{\EV}_{\text{\textsc{m}}}}    
\newcommand{\DVM}{{\DV}_{\text{\textsc{m}}}}    
\newcommand{\HVM}{{\HV}_{\text{\textsc{m}}}}    

\newcommand{\AMM}{{\AV}_{\text{\textsc{mm}}}}    
\newcommand{\BMM}{{\BV}_{\text{\textsc{mm}}}}    
\newcommand{\EMM}{{\EV}_{\text{\textsc{mm}}}}    
\newcommand{\phiMM}{{\phi}_{\text{\textsc{mm}}}} 

\newcommand{\AML}{{\AV}_{\text{\textsc{ml}}}}    
\newcommand{\BML}{{\BV}_{\text{\textsc{ml}}}}    
\newcommand{\EML}{{\EV}_{\text{\textsc{ml}}}}    
\newcommand{\DML}{{\DV}_{\text{\textsc{ml}}}}    
\newcommand{\phiML}{{\phi}_{\text{\textsc{ml}}}} 

\newcommand{\AMBI}{{\AV}^{\text{\textrm{sf}}}_{\text{\textsc{mbi}}}} 
\newcommand{\BMBI}{{\BV}^{\text{\textrm{sf}}}_{\text{\textsc{mbi}}}} 
\newcommand{\EMBI}{{\EV}^{\text{\textrm{sf}}}_{\text{\textsc{mbi}}}} 
\newcommand{\DMBI}{{\DV}^{\text{\textrm{sf}}}_{\text{\textsc{mbi}}}} 
\newcommand{\phiMBI}{{\phi}^{\text{\textrm{sf}}}_{\text{\textsc{mbi}}}} 

\newcommand{\AMBIpt}{{\AV}_{\text{\textsc{mbi}}}} 
\newcommand{\BMBIpt}{{\BV}_{\text{\textsc{mbi}}}} 
\newcommand{\EMBIpt}{{\EV}_{\text{\textsc{mbi}}}} 
\newcommand{\DMBIpt}{{\DV}_{\text{\textsc{mbi}}}} 
\newcommand{\HMBIpt}{{\HV}_{\text{\textsc{mbi}}}} 
\newcommand{\phiMBIpt}{{\phi}_{\text{\textsc{mbi}}}} 
\newcommand{\TMBIpt}{{T}_{\text{\textsc{mbi}}}} 
\newcommand{\ThetaMBIpt}{{\Theta}_{\text{\textsc{mbi}}}} 

\newcommand{\BMBIptreg}{{\BV}_{\text{\textsc{mbi}}}^{\text{\textrm{reg}}}}  
\newcommand{\EMBIptreg}{{\EV}_{\text{\textsc{mbi}}}^{\text{\textrm{reg}}}}  

\newcommand{\RHJ}{{R}_{\text{\textsc{hj}}}} 
\newcommand{\SHJ}{{S}_{\text{\textsc{hj}}}} 

\newcommand{\RQM}{{R}_{\text{\textsc{qm}}}} 
\newcommand{\SQM}{{S}_{\text{\textsc{qm}}}} 

\newcommand{\nullV}{\vect{O}}
\newcommand{\AV}{\pmb{{\cal A}}}
\newcommand{\BV}{\pmb{{\cal B}}}
\newcommand{\DV}{\pmb{{\cal D}}}
\newcommand{\EV}{\pmb{{\cal E}}}

\newcommand{\HV}{\pmb{{\cal H}}}

\newcommand{\LV}{\vect{L}}
\newcommand{\PV}{\vect{P}}
\newcommand{\PiV}{\vect{\Pi}}

\newcommand{\LVf}{{\LV}_{\text{\textrm{f}}}}    
\newcommand{\PVf}{{\PV}_{\text{\textrm{f}}}}    
\newcommand{\Ef}{{E}_{\text{\textrm{f}}}}       
\newcommand{\mf}{{m}_{\text{\textrm{f}}}}       

\newcommand{\nablaq}{\nabla_{\!\qV}}

\newcommand{\nab}{\vect{\nabla}}

\newcommand{\alphaQ}{\vect{\alpha}}
\newcommand{\betaQ}{{\beta}}

\newcommand{\qdot}[1]{{\stackrel{\,\circ}{{#1}}}}
\newcommand{\qddot}[1]{{\stackrel{\circ\circ}{{#1}}}}

\newcommand{\equiveg}{\stackrel{ _\sim}{=}}

\renewcommand{\geq}{\geqslant}

\newcommand{\cE}{{\cal E}} 
\newcommand{\cL}{{\cal L}} 
\newcommand{\cP}{{\cal P}} 

\reversemarginpar

\numberwithin{equation}{section}

\begin{document}


\title{\uppercase{On the motion of point defects \\
                  in relativistic fields}}

\author{\textbf{Michael K.-H. KIESSLING}\\
                Department of Mathematics, Rutgers University\\
                110 Frelinghuysen Rd., Piscataway, NJ 08854, USA\\ \\ \\
\textrm{Version of July 07, 2011.}\\
\textrm{This printout was typeset with \LaTeX\ on}}
\maketitle
\thispagestyle{empty}

\begin{abstract}
{\noindent 
We inquire into classical and quantum laws of motion for the point charge sources 
in the nonlinear Maxwell--Born--Infeld field equations of classical electromagnetism
in flat and curved Einstein spacetimes.
}
\end{abstract}

\vfill
\hrule
\smallskip
To appear (with different layout) in the proceedings of the conference

``Quantum Field Theory and Gravity'',  Regensburg 2010 
(Felix Finster, 

Olaf M\"uller, Marc Nardmann, J\"urgen Tolksdorf, Eberhard Zeidler, eds.).
\smallskip

\copyright(2011) \small{The author. Reproduction of this preprint, in its entirety, is 

permitted for non-commercial purposes only.}

\newpage



                \section{Introduction}

{\small{\hskip2truecm \hfill Shortly after it was realized that there was a problem, quantum 

        \hskip2.2truecm \   physics was invented. Since then we have been doing quantum \ \ 

	\hskip1.9truecm \ \ { }physics, and the problem was forgotten. But it is still with us!\hfill

\hfill\hfill Detlef D\"urr and Sergio Albeverio, late 1980s}}
\medskip

\noindent
  The ``forgotten problem'' that D\"urr and Albeverio were talking about some 20+ years ago 
is the construction of a consistent classical theory for the joint evolution of 
electromagnetic fields and their point charge sources.
  Of course the problem was not completely forgotten, but it certainly has become a backwater 
of mainstream physics with its fundamental focus on quantum theory: first quantum field theory 
and quantum gravity, then  string, and in recent years now $M$-theory.
  Unfortunately, more than a century of research into quantum physics has not yet produced 
a consistent quantum field theory of the electromagnetic interactions without artificial 
irremovable mathematical regularizers; the incorporation of the weak and strong interactions in
the standard model has not improved on this deficiency. 
 The consistent theory of quantum gravity has proved even more elusive, and nobody (presumably)
knows whether $M$-theory is ever going to see the light of the day.
  So it may yet turn out that the ``forgotten classical problem'' will be solved first. 

  In the following, I will report on some exciting recent developments towards the solution of 
the ``forgotten classical problem'' in general-relativistic spacetimes, in terms of the coupling 
of the Einstein--Maxwell--Born--Infeld theory for an electromagnetic spacetime with point defects 
(caused by point charge sources) to a Hamilton-Jacobi theory of motion for these point defects.
  Mostly I will talk about the special-relativistic spacetime limit, though.
  Since I want to emphasize the evolutionary aspects of the theory, I will work in a space+time
splitting of spacetime rather than using the compact formalism of spacetime geometry.
  Furthermore I will argue that this putative solution to the classical problem also teaches us 
something new about the elusive consistent quantum theory of electromagnetism with point sources, 
and its coupling to gravity.
  Namely, while the spacetime structure and the electromagnetic fields will still be treated at 
the classical level, replacing our classical Hamilton-Jacobi law of motion for the electromagnetic 
point defects by a de Broglie--Bohm--Dirac quantum law of motion for the point defects 
yields a reasonable ``first quantization with spin'' of the classical theory of motion of
point defects in the fields, which has the additional advantage that it doesn't suffer from 
the infamous measurement problem. 
  In all of these approaches, the structure of spacetime is classical. 
  I will have to leave comments on the pursuit of the photon and
the graviton, and quantum spacetimes to a future contribution.

   I now begin by recalling the ``forgotten classical problem.'' 

                \section{\hskip-2.5pt Lorentz electrodynamics with point charges}
  
	I briefly explain why the formal equations of classical Lorentz electrodynamics with 
point charges fail to yield a well-defined classical theory of electromagnetism.\footnote{I
        	hope that this also dispels the perennial myth in the plasma physics literature that 
		these ill-defined equations were ``the fundamental equations of a classical plasma.''}

                \subsection{Maxwell's field equations}

	I prepare the stage by recalling Maxwell's field equations of electromagnetism 
in Minkowski spacetime, written with respect to any convenient flat foliation 
(a.k.a. Lorentz frame) into space points $\sV\in\Rset^3$ at time $t\in\Rset$.
	Suppose a relativistic theory of matter has supplied an electric charge density $\rho (t,\sV)$ 
and an electric current vector-density $\jV (t,\sV)$, satisfying
the local law of charge conservation,
\begin{equation}
\textstyle
        \pddt\rho (t,\sV) + \nab\cdot\jV(t,\sV)  
=\label{eq:MrhojLAW}
	0.
\end{equation}
\emph{Maxwell's electromagnetic field equations} comprise the two evolution equations
\begin{alignat}{1}
\textstyle
   \frac{1}{c}
\pddt{\BVM(t,\sV)}
&= \label{eq:MdotB}
        - \nab\times\EVM(t,\sV) \, ,
\\
\textstyle
        \frac{1}{c}\pddt{\DVM(t,\sV)}
&= 
        + \nab\times\HVM(t,\sV)  - 4\pi \textstyle{\frac{1}{c}} \jV(t,\sV) \, ,
\label{eq:MdotD}
\end{alignat}
and the two constraint equations 
\begin{alignat}{1}
        \nab\cdot \BVM(t,\sV)  
&= \label{eq:MdivB}
        0\, ,
\\
        \nab\cdot\DVM(t,\sV)  
&=
        4 \pi \rho (t,\sV) \, .
\label{eq:MdivD}
\end{alignat}
	These field equations need to be supplemented by a relativistic ``constitutive law'' which 
expresses the electric and magnetic fields $\EVM$ and $\HVM$ in terms of the magnetic induction 
field $\BVM$ and the electric displacement field $\DVM$.
	The constitutive law reflects the ``constitution of matter'' 
and would have to be supplied by the theory of matter carrying $\rho$ and $\jV$.
	(Later we will adopt a different point of view.)

	For matter-free space Maxwell proposed
\begin{alignat}{1}
        \HVM(t,\sV)  
&= \label{eq:MlawBisH}
        \BVM(t,\sV)  \, ,
\\
        \EVM(t,\sV)  
&=
        \DVM(t,\sV) \, .
\label{eq:MlawEisD}
\end{alignat}
	The system of Maxwell field equations 
\refeq{eq:MdotB}, \refeq{eq:MdotD}, \refeq{eq:MdivB}, \refeq{eq:MdivD} 
with $\rho \equiv 0$ and $\jV\equiv \vect{0}$, supplemented by 
Maxwell's ``law of the pure aether'' \refeq{eq:MlawBisH} and \refeq{eq:MlawEisD}, will 
be called the \emph{Maxwell--Maxwell field equations}.
	They feature a large number of conserved quantities \cite{AncoThe}, 
including the field energy, the field momentum, 
\newpage

\noindent
and the field angular momentum, given by, respectively (cf. \cite{AbrahamBOOK}, \cite{JacksonBOOKb}),
\begin{alignat}{1}
  \Ef
&= \label{eq:FIELDenergyMM}
  \frac{1}{8\pi}\int_{\mathbb R^3}\bigl(|\EMM(t,\sV)|^2 +|\BMM(t,\sV)|^2\bigr)\,\drm^3s,
\\
  \PVf
&= \label{eq:FIELDimpulsMM}
  \frac{1}{4\pi c} \int_{\mathbb R^3} \EMM(t,\sV)\times\BMM(t,\sV) \, \drm^3s\, , 
\\
  \LVf
&= \label{eq:FIELDdrehimpulsMM}
  \frac{1}{4\pi c} \int_{\mathbb R^3} \sV\times(\EMM(t,\sV)\times\BMM(t,\sV)) \,\drm^3s\, .
\end{alignat}
	These integrals will retain their meanings also in the presence of point sources.

           \subsection{The Maxwell--Lorentz field equations}

	Although Maxwell pondered atomism --- think of Maxwell's velocity 
distribution and the Maxwell--Boltzmann equation in the kinetic theory of gases ---,
it seems that he did not try to implement atomistic notions of matter into 
his electromagnetic field equations. 
	This step had to wait until the electron was discovered, by Wiechert \cite{WiechertsEXPePAP} 
and Thomson \cite{jjthomsonB} (see \cite{pippard} and  \cite{Jost}).
	Assuming the electron to be a point particle with charge $-e$, the Maxwell field equations 
for a single electron embedded in Maxwell's ``pure aether'' at $\QV(t)\in\Rset^3$ at time $t$
become the \emph{Maxwell--Lorentz field equations} (for a single electron),
\begin{alignat}{1}
\textstyle
        \frac{1}{c}\pddt{\BML(t,\sV)}
&= \label{eq:MLdotB}
        -\nab\times\EML(t,\sV) \, ,
\\
\textstyle
         \frac{1}{c}\pddt{\EML(t,\sV)}
&= \label{eq:MLdotE}
        +\nab\times\BML(t,\sV)  + 4\pi  e\delta_{\QV(t)}(\sV) \textstyle{\frac{1}{c}} \dot{\QV}(t)\, ,
\\
        \nab\cdot \BML(t,\sV)  
&= \label{eq:MLdivB}
        0\, ,
\\
        \nab\cdot\EML(t,\sV)  
&=\label{eq:MLdivE}
       - 4 \pi e \delta_{\QV(t)} (\sV) \, ,
\end{alignat}
where ``$\delta(\,\cdot\,)$'' is Dirac's delta function, and $\dot\QV(t)$ 
the velocity of the point electron. 
	Note that the point charge ``density'' $\rho(t,\sV)\equiv -e \delta_{\QV(t)}(\sV)$ and 
current vector-``density'' $\jV(t,\sV)\equiv - e\delta_{\QV(t)}(\sV)\dot{\QV}(t)$ jointly
satisfy the continuity equation \refeq{eq:MrhojLAW} in the sense of distributions; of 
course, the Maxwell--Lorentz field equations have to be interpreted in the sense 
of distributions, too.
	As is well-known, the Maxwell--Lorentz field equations are covariant under the 
Poincar\'e group; of course, the position and velocity of all the point charges
are transformed accordingly as well. 

	Given any twice continuously differentiable, subluminal ($|\dot\QV(t)|<c$) motion
$t\mapsto\QV(t)$, the Maxwell--Lorentz field equations are solved by 

\vskip-.6truecm
\begin{alignat}{1}
\hskip-.8truecm
        \ELWr(t,\sV)  
&=\label{eq:LWsolE}
-e {\textstyle{\frac{1}{(1-\nV\cdot\dot\QV/c)^3}}}\Bigl(\frac{\nV-\dot\QV/c}{\gamma^2 r^2}
 + \frac{\nV\times[(\nV-\dot\QV/c)\times\ddot\QV/c^2]}{r}\Bigr)\Big|_{\mathrm{ret}}
\\
\hskip-.8truecm
        \BLWr(t,\sV)  
&= \label{eq:LWsolB}
        \nV|_{_{\mathrm{ret}}}\times \ELWr(t,\sV)
\, ,
\end{alignat}
where $\nV = (\sV-\QV(t))/r$ with $r=|\sV-\QV(t)|$,  $\gamma^2 = 1/({1-|\dot\QV(t)|^2/c^2})$,
and where ``${\mathrm{ret}}$'' means that
the $t$-dependent functions $\QV(t)$, $\dot\QV(t)$, and $\ddot\QV(t)$ are to be
\newpage

\noindent
evaluated at the retarded time $t_{\mathrm{ret}}$ defined implicitly by 
$c(t-t_{\mathrm{ret}}) = |\sV-\QV(t_{\mathrm{ret}})|$; 
see \cite{lienard}, \cite{Wiechert}.
 By linearity, the general solution of the Maxwell--Lorentz equations with many
point sources is now obtained by adding all their pertinent Li\'enard--Wiechert fields 
to the general solution to the Maxwell--Maxwell equations.

  Since the Maxwell--Lorentz field equations are consistent with \emph{any} smooth subluminal 
motion $t\mapsto \QV(t)$, to determine the physical motions a \emph{law of motion} for the 
point electron has to be supplied.
  For the physicist of the late 19th and early 20th centuries this meant a Newtonian law of motion.

           \subsection{The Lorentz force}
           \subsubsection{The test charge approximation: all is well!}

  Lorentz \cite{lorentzENCYCLOP}, Poincar\'e \cite{poincare}, and  Einstein \cite{einsteinA}
showed that at the level of the \emph{test particle} approximation,
Newton's law for the rate of change of its mechanical momentum, equipped with the 
Lorentz force \cite{lorentzFORCE}
\begin{equation}
  \frac{\drm \PV(t)}{\drm t}
= \label{eq:NLeOFm}
  -e\Bigl[\EV(t,\QV(t)) + \textstyle{\frac{1}{c}}\dot\QV(t) \times\BV(t,\QV(t))\Bigr],
\end{equation}
combined with the relativistic law between mechanical momentum and velocity,
\begin{equation}
  \frac{1}{c} \frac{\drm \QV(t)}{\drm t}
= \label{eq:ELPpv}
  \frac{\PV(t)}{\sqrt{\me^{2}c^2 + |\PV(t)|^2}} ,
\end{equation}
provides an empirically highly accurate law of motion for the point electron;
in \refeq{eq:NLeOFm}, $\EV=\EMM$ and $\BV=\BMM$ are solutions of the Maxwell--Maxwell 
field equations, and the $\me$ in \refeq{eq:ELPpv} is the empirical inert rest mass of 
the physical electron, which is \emph{defined} through this test particle law!   
	This law of test particle motion is globally well-posed as a Cauchy problem for the map
$t\mapsto(\PV(t),\QV(t))$ \emph{given} any Lipschitz continuous so-called \emph{external fields}
$\EMM$ and $\BMM$.

           \subsubsection{The Lorentz self-force: infinite in all directions!}

	Clearly, fundamentally there is no such thing as a ``test charge'' in ``external fields;'' only
the total fields, the solutions to the Maxwell--Lorentz field equations with the point charge as 
source, can be fundamental in this theory.
	However, the law of motion \refeq{eq:NLeOFm}, \refeq{eq:ELPpv} is \emph{a priori undefined} when 
{formally} coupled with \refeq{eq:MLdotB}, \refeq{eq:MLdotE}, \refeq{eq:MLdivB}, \refeq{eq:MLdivE},
so that $\EV = \EML \equiv \ELWr+ \EMM$ and $\BV= \BML\equiv \BLWr+\BMM$ in \refeq{eq:NLeOFm};
indeed, inspection of \refeq{eq:LWsolE}, \refeq{eq:LWsolB} makes it
plain that ``$\ELWr(t,\QV(t))$'' and ``$\BLWr(t,\QV(t))$'' are ``infinite in all directions,''
by which I mean that for any limit $\sV\to\QV(t)$ of $\ELWr(t,\sV)$ and $\BLWr(t,\sV)$,
the field magnitudes diverge to infinity while their limiting directions, whenever such exist, 
depend on how the limit is taken.

	Lorentz and his peers interpreted the infinities to mean that the electron is not 
really a point particle.
	To uncover its structure by computing details of the motion which depend on
its structure became the goal of ``classical electron theory''
\cite{lorentzBOOKa,WiechertsBIGpaper,Abraham,lorentzENCYCLOP,AbrahamBOOK,lorentzBOOKb}.
	This story is interesting in its own right, see \cite{rohrlichBOOK,yaghjianBOOK}, 
but would lead us too far astray from our pursuit of a well-defined theory of electromagnetism 
with point charges.\footnote{However, I do take the opportunity
		to advertise a little-known but very important fact. 
	By postulating energy-momentum (and angular momentum) conservation of the fields
		alone, Abraham and Lorentz derived an effective Newtonian test particle law of motion 
		from their ``purely electromagnetic models'' in which both the external Lorentz force
		\emph{and the particle's inertial mass} emerged through an expansion in a small parameter.

	Yet, in \cite{AppKieAOP} Appel and myself showed that the Abraham--Lorentz proposal
		is mathematically \emph{inconsistent}: generically their ``fundamental law of 
		motion'' does not admit a solution at all!
	But how could Abraham and Lorentz arrive at the correct approximate law of 
		motion in the leading order of their expansion? 
	The answer is: if you take an inconsistent nonlinear equation and \emph{assume} that it has 
		a solution which provides a small parameter, then formally expand the
		equation and its hypothetical solution in a power series w.r.t. this parameter, then 
		truncate the expansion and treat the retained expansion coefficients as free 
		parameters to be matched to empirical data, \emph{then} you may very well end up 
		with an accurate equation --- all the inconsistencies are hidden in the pruned-off 
		part of the expansion!
	But you haven't \emph{derived} anything, seriously.}

           \subsubsection{Regularization and renormalization?}

	Precisely because ``$\ELWr(t,\QV(t))$'' and ``$\BLWr(t,\QV(t))$'' are ``infinite 
\emph{in all directions},'' it is tempting to inquire into the possibility 
of \emph{defining} r.h.s.\refeq{eq:NLeOFm} for solutions of the Maxwell--Lorentz 
field equations through a point limit 
$\varrho(\sV|\QV(t))\to\delta_{\QV(t)}(\sV)$ 
of a Lorentz force field 
$\EML(t,\sV) + {\scriptstyle{\frac{1}{c}}}\dot\QV(t) \times\BML(t,\sV)$ 
averaged over some normalized regularizing density $\varrho(\sV|\QV(t))$. 
	Of course, for the Maxwell--Maxwell part of the Maxwell--Lorentz fields this 
procedure yields 
$\EMM(t,\QV(t)) + {\scriptstyle{\frac{1}{c}}}\dot\QV(t) \times\BMM(t,\QV(t))$.
	For the Li\'enard--Wiechert fields on the other hand such a definition via 
regularization cannot be expected to be unique (if it leads to a finite result at all);
indeed, one should expect that one can obtain \emph{any} limiting averaged force vector
by choosing a suitable family of averages around the location of the point charge, so 
that such a definition of the Lorentz force is quite arbitrary.
	In the best case one could hope to find some physical principle which selects a 
unique way of averaging the Lorentz force field and of taking the point limit of it.
	Meanwhile, in the absence of any such principle, one may argue that anything else 
but taking the limit $R\to 0$ of a uniform average of 
$\ELWr(t,\sV)+{\scriptstyle{\frac{1}{c}}}\dot\QV(t)\times\BLWr(t,\sV)$ 
over a sphere of radius $R$ centered at $\QV(t)$ \emph{in the instantaneous rest 
frame} of the point charge, followed by a boost back into the original Lorentz frame,
would be \emph{perverse}.
\newpage

	Unfortunately, work by Dirac \cite{DiracA} has revealed that the point limit of such 
a ``spherical averaging'' does \emph{not} produce  a finite r.h.s.\refeq{eq:NLeOFm} 
with $\ELWr+\EMM$ and $\BLWr+\BMM$ in place of $\EV$ and $\BV$.
	Instead, to leading order in powers of $R$ the spherical average of the Lorentz ``self-force'' field 
$\ELWr(t,\sV)+{\scriptstyle{\frac{1}{c}}}\dot\QV(t)\times\BLWr(t,\sV)$ is given 
by\footnote{The fact that $e^2/2R$ coincides with the electrostatic field energy of a 
		surface-charged sphere of radius $R$ is a consequence of Newton's theorem. 
	The regularization of the Lorentz force field of a point charge	by ``spherical 
		averaging'' should \emph{not} be confused with setting up a dynamical Lorentz 
		model of a ``surface-charged sphere,'' e.g. \cite{AppKieAOP}, \cite{AppKieLMP}.} 
$-\frac{e^2}{2c^2}R^{-1}\pddt(\gamma(t)\dot\QV(t))$.
  Unless the acceleration vanishes at time $t$, this term diverges 
$\uparrow\infty$ in magnitude when $R\downarrow 0$.
	Incidentally, the fact that this term vanishes when the point particle is 
unaccelerated shows that the infinities of the Lorentz self-force in
stationary motion have been removed by spherical averaging.

        In addition to the divergence problems of the Lorentz self-force on a point charge,
the field energy integral \refeq{eq:FIELDenergyMM} diverges for the 
Li\'enard--Wiechert fields because of their local singularity at $\sV=\QV(t)$ 
(a ``classical UV divergence'').\footnote{The field energy integral 
	\refeq{eq:FIELDenergyMM} for the Li\'enard--Wiechert fields diverges 
	also because of their slow decay as $\sV\to\vect{\infty}$, but this 
	``classical IR divergence'' can be avoided by adding a suitable solution 
	of the Maxwell--Maxwell field equations.}
  This confronts us also with the problem that by Einstein's $E=mc^2$ \cite{einsteinB}
the electromagnetic field of a point charge should attach an infinite inert mass to 
the point charge.
	In a sense, this is precisely what the in leading order divergent Lorentz self-force
term ``$\lim_{R\downarrow 0}-\frac{e^2}{2c^2}R^{-1}\pddt(\gamma(t)\dot\QV(t))$'' 
expresses.
	But how does that fit in with the finite $\me$ in \refeq{eq:ELPpv}?
	There is no easy way out of this dilemma.

	In \cite{DiracA} Dirac proposed to assign an $R$-dependent bare 
mass $\mbare(R)$ to the averaging sphere of radius $R$, such that 
$\mbare(R) +\frac{e^2}{2c^2}R^{-1}\to \me$ as $R\downarrow 0$, 
where $\me$ is the empirical rest mass of the electron (see above).
	It seems that this was the first time such a radical step was proposed, a precursor 
to what eventually became ``renormalization theory;'' cf. \cite{gitteletal}.
	Dirac's procedure in effect removes the divergent ``self-force'' from 
r.h.s.\refeq{eq:NLeOFm}, with $\ELWr+\EMM$ and $\BLWr+\BMM$ in place of $\EV$ and 
$\BV$, and produces the so-called Abraham--Lorentz--Dirac equation, 
viz.\footnote{The complicated correction term to the 
		``external Lorentz force'' at r.h.s.\refeq{eq:ALDeOFm} is 
		simply the space part of the Laue \cite{vonlaue} four-vector
		$({2e^2}/{3c^3}) (\gQ +\uQ\tens\uQ)\cdot{\qddot\uQ}$, divided by $\gamma(t)$.
	Here, $\uQ$ is the dimensionless four-velocity of the point charge,
		$\qdot{ }$ means derivative w.r.t. ``$c\times$proper time,''
		and $\gQ$ is the metric tensor for Minkowski spacetime with signature $(-,+,+,+)$.
	Note that $\gQ +\uQ\tens\uQ$ projects onto the subspace four-orthogonal to ${\uQ}$.}
\begin{alignat}{1}
  \frac{\drm \PV(t)}{\drm t}
= \label{eq:ALDeOFm}
  &-e\Bigl[\EMM(t,\QV(t)) + \textstyle{\frac{1}{c}}\dot\QV(t) \times\BMM(t,\QV(t))\Bigr]\\
\notag
  &+ \textstyle{\frac{2e^2}{3c^3}}
\left[\II + \gamma^2 \frac{1}{c^2}\dot\QV\otimes\dot\QV\right]
\!\cdot\!
\left[3\gamma^4 \frac{1}{c^2}(\dot\QV\cdot\ddot\QV)\ddot\QV +\gamma^2\dddot\QV\right](t)
\notag
\end{alignat}
coupled with \refeq{eq:ELPpv}; here $\II$ is the identity operator.
	The occurrence in \refeq{eq:ALDeOFm} of the third 
time derivative of $\QV(t)$ signals that our troubles are not over yet.
	Viewed as a Cauchy problem, $\ddot\QV(0)$ has now to be prescribed in addition to
the familiar initial data $\QV(0)$ and $\dot\QV(0)$, and most choices will lead to 
self-accelerating run-away solutions.
	To eliminate these pathological solutions amongst all solutions with the 
same initial data $\QV(0)$ and $\dot\QV(0)$, Dirac \cite{DiracA} integrated 
\refeq{eq:ALDeOFm} once with the help of an integrating factor, obtaining an integro-differential 
equation which is of second order in the time derivative of $\QV(t)$ but now with the applied force 
integrated over the whole future of the trajectory.
	This implies that the remaining solutions are ``pre-accelerated,''\footnote{``It is to be 
	hoped that some day the real solution of the problem of the charge-field 
	interaction will look different, and the equations describing nature will 
	not be so highly unstable that the balancing act can only succeed by having 
	the system correctly prepared ahead of time by a convenient coincidence.''
	Walter Thirring, p. 379 in \cite{ThirringBOOKa}.}
and they can be non-unique \cite{caratigalganietal}.

	Note that Dirac's ad hoc procedure actually means a step in the direction of taking a
renormalized point-particle limit in a family of ``extended electron''  models, though 
Dirac made no attempt (at this point) to come up with a consistent dynamical model 
of an extended electron.
	If he had, he would have found that for a family of dynamically consistent models 
of a ``relativistically spinning Lorentz sphere'' one cannot take the renormalized 
point charge limit, cf. \cite{AppKieAOP}.
 
    \subsection{\hskip-.3truecm The effective equation of motion of Landau--Lifshitz}

	Despite the conceptual problems with its third-order derivative, the Abraham-Lorentz--Dirac 
equation has served Landau and Lifshitz \cite{landaulifshitzBOOK} (and Peierls) as point of departure to 
arrive at an effective second-order equation of motion which is free of runaway solutions and pre-acceleration.
	They argued that whenever the familiar second-order equation of test charge motion is 
empirically successful, the second line at r.h.s.\refeq{eq:ALDeOFm} should be treated 
as a \emph{tiny} correction term to its first line.
	As a rule of thumb this should be true for test particle motions with trajectories whose
curvature $\kappa$ is much smaller than ${\me c^2}/{e^2}$, the reciprocal of the ``classical electron radius.'' 
	But then, in leading order of a formal expansion in powers of the small parameter $\kappa {e^2}/{\me c^2}$, 
the third time derivative of $\QV(t)$ in \refeq{eq:ALDeOFm} can be expressed in terms of $\QV(t)$ and $\dot\QV(t)$, 
obtained from the equations of test particle motion \refeq{eq:NLeOFm} and \refeq{eq:ELPpv} 
as follows: invert \refeq{eq:ELPpv} to obtain $\PV(t) = \me\gamma(t)\dot\QV(t)$, then take the 
time derivative of this equation and solve for $\ddot\QV(t)$ to get
\begin{alignat}{1}
\ddot\QV(t) 
= \label{eq:QddOfPd}
\textstyle{\frac{1}{\me\gamma(t)}}
\left[\II -  \frac{1}{c^2}\dot\QV(t)\otimes\dot\QV(t)\right]\!\cdot\!\dot\PV(t).
\end{alignat}

\vskip-0.2truecm
\noindent
     Now substitute the first line at r.h.s.\refeq{eq:ALDeOFm} for $\dot\PV(t)$ at r.h.s.\refeq{eq:QddOfPd},
then take another time derivative of the so rewritten equation \refeq{eq:QddOfPd} to obtain
$\dddot\QV(t)$ in terms of $\QV,\dot\QV,\ddot\QV$; for the $\ddot\QV$ terms in this expression, now
resubstitute \refeq{eq:QddOfPd}, with the first line at r.h.s.\refeq{eq:ALDeOFm} substituted for $\dot\PV(t)$,
to get $\dddot\QV(t)$ in terms of $\QV,\dot\QV$.
	Inserting this final expression for $\dddot\QV(t)$ into the second line at r.h.s.\refeq{eq:ALDeOFm} 
yields the so-called Landau--Lifshitz equation of motion of the physical electron,\footnote{This tedious calculation
		becomes simplified in the four-vector formulation mentioned in footnote 5.
	The Landau--Lifshitz reasoning for the leading order $\qddot\uQ$ yields 
		the proper time derivative of the external Lorentz--Minkowski force, divided by $\me$,
		in which in turn the external Lorentz--Minkowski force, divided by $\me$,
		is resubstituted for the $\qdot\uQ$ term.
	Projection onto the subspace orthogonal to $\uQ$ and multiplication by 
		$\frac{2e^2}{3c^3}$ yields the Landau--Lifshitz approximation to the Laue four-vector. 
	Its space part divided by $\gamma(t)$ gives the pertinent approximation to the second
		line at r.h.s.\refeq{eq:ALDeOFm} in terms of $\QV(t)$ and $\dot\QV(t)$.} 
coupled with \refeq{eq:ELPpv}.
	The error made by doing so is of higher order in the small expansion parameter than the retained terms.
	I refrain from displaying the Landau--Lifshitz equation because its intimidating
appearance does not add anything illuminating here.

	Of course, the combination of Dirac's ad hoc renormalization procedure with the 
heuristic Landau--Lifshitz approximation step does not qualify as a satisfactory 
``derivation'' of the Landau--Lifshitz equation of motion from ``first principles.'' 
	A rigorous derivation from an extended particle model has been given by Spohn and 
collaborators, cf.  \cite{spohnBOOK}.
	This derivation establishes the status of the Landau--Lifshitz equation as an 
effective equation of motion for the geometrical center of a particle with structure, 
not the point particle hoped for by Dirac.
	Whether it will ever have a higher (say, at least asymptotic) status for a proper 
theory of charged point-particle motion remains to be seen.
	In any event, this equation seems to yield satisfactory results for many practical purposes. 

           \section{Wheeler--Feynman electrodynamics}

	Before I can move on to the next stage in our quest for a proper theory of motion of 
the point charge sources of the electromagnetic fields, I need to mention the radical 
solution to this classical problem proposed by Fokker, Schwarzschild, and Tetrode 
(see \cite{barutBOOKa}, \cite{spohnBOOK}) and its amplification by Wheeler--Feynman \cite{Wheelerfeynman}.
	This ``action-at-a-distance'' theory takes over from formal Lorentz electrodynamics the Li\'enard--Wiechert 
fields associated to each point charge, but there are no additional external Maxwell--Maxwell fields.
	Most importantly, a point charge is not directly affected by its own Li\'enard--Wiechert field. 
	Therefore the main problem of formal Lorentz electrodynamics, the self-interaction 
of a point charge with its own  Li\'enard--Wiechert field, simply does not exist!

	In the Fokker--Schwarzschild--Tetrode theory, the law of motion of an electron is 
given by equations \refeq{eq:NLeOFm}, \refeq{eq:ELPpv}, though with $\EV$ and $\BV$ 
now standing for the arithmetic means $\frac{1}{2}(\ELWr + \ELWa)$ and 
$\frac{1}{2}(\BLWr +\BLWa)$ of the \emph{retarded and advanced} Li\'enard--Wiechert 
fields summed over \emph{all the other} point electrons.
	Nontrivial motions can occur only in the many-particle Fokker--Schwarzschild--Tetrode theory.
  
	While the Fokker--Schwarzschild--Tetrode electrodynamics does not suffer from the 
infinities of formal Lorentz electrodynamics, it raises another formidable problem: 
its second-order equations of motion for the system of point charges do not pose 
a Cauchy problem for the traditional classical state variables $\QV(t)$ and 
$\dot\QV(t)$ of these point charges.
	Instead, given the classical state variables $\QV(t_0)$ and $\dot\QV(t_0)$ of each 
electron at time $t_0$, the computation of their accelerations at time $t_0$ 
requires the knowledge of the states of motion of all point electrons at infinitely 
many instances in the past \emph{and in the future}. 
	While it would be conceivable in principle, though certainly not possible in 
practice, to find some historical records of all those past events, how could we 
anticipate the future without computing it from knowing the present and the past? 

	Interestingly enough, though, Wheeler and Feynman showed that the 
Fokker--Schwarzschild--Tetrode equations of motion can be recast as  
Abraham--Lorentz--Dirac equations of motion for each point charge, 
though with the external Maxwell--Maxwell fields replaced by the retarded 
Li\'enard--Wiechert fields of all the other point charges, \emph{provided}
the \emph{Wheeler--Feynman absorber identity} is valid.
	While this is still not a Cauchy problem for the classical state variables 
$\QV(t)$ and $\dot\QV(t)$ of each electron, at least one does not need to anticipate 
the future anymore.
  
	The Fokker--Schwarzschild--Tetrode and Wheeler--Feynman theories 
are mathematically fascinating in their own right, but pose very difficult problems.
	Rigorous studies have only recently begun \cite{bauer}, \cite{BauerDeckertDuerr}, \cite{Deckert}.
  
                \section{Nonlinear electromagnetic field equations}

	Gustav Mie \cite{MieFELDTHEORIE} worked out the special-relativistic framework for 
fundamentally nonlinear electromagnetic field equations without point charges 
(for a modern treatment, see \cite{christodoulou}).
	Twenty years later Mie's work became the basis for Max Born's assault on the 
infinite self-energy problem of a point charge.

\vskip-5pt
                \subsection{Nonlinear self-regularization}
 Born \cite{BornA} argued that the dilemma of the infinite electromagnetic self-energy 
of a point charge in formal Lorentz electrodynamics is caused by
Maxwell's ``law of the pure aether,''\footnote{After its demolition by Einstein, 
	``aether'' will be recycled here as shorthand for what physicists call ``electromagnetic vacuum.''}
which he suspected to be valid only asymptotically, viz.
${\EVM} \sim {\DVM}$ and ${\HVM} \sim{\BVM}$ in the weak field limit.
 In the vicinity of a point charge, on the other hand, where the Coulombic $\DML$ 
field diverges in magnitude as $1/r^2$, nonlinear deviations of the true law of the 
``pure aether'' from Maxwell's law would become significant and ultimately remove the 
infinite self-field-energy problems.
  The sought-after nonlinear ``aether law'' has to satisfy the requirements that the
resulting electromagnetic field equations derive from a Lagrangian which:
\smallskip

(P) is covariant under the Poincar\'e group;

(W) is covariant under the Weyl (gauge) group;

(M) reduces to the Maxwell--Maxwell Lagrangian in the weak field limit;

(E) yields finite field-energy solutions with point charge sources.

\smallskip
\noindent
  The good news is that there are ``aether laws'' formally satisfying criteria 
(P), (W), (M), (E).
  The bad news is that there are too many ``aether laws'' which satisfy criteria 
(P), (W), (M), (E)
formally, so that additional requirements are needed.

  Since nonlinear field equations tend to have solutions which form singularities in 
finite time (think of shock formation in compressible fluid flows), it is reasonable 
to look for the putatively least troublesome nonlinearity. 
 In 1970, Boillat \cite{Boillat}, and independently Plebanski \cite{Plebanski}, 
discovered that adding to (P),(W),(M),(E) the requirement that the electromagnetic 
field equations:
\smallskip

(D) are linearly completely degenerate,

\smallskip
\noindent
a \emph{unique} one-parameter family of field equations emerges, indeed the one 
proposed by Born and Infeld \cite{BornInfeldBa,BornInfeldBb}\footnote{While the unique 
		characterization of the Maxwell--Born--Infeld field equations 
		in terms of (P),(W),(M),(E),(D) was apparently not known to Born 
		and his contemporaries, the fact that these field equations satisfy, 
		beside (P),(W),(M),(E), also (D) is mentioned in passing already on 
		p.102 in \cite{ErwinDUBLINa} as the absence of birefringence (double 
		refraction), meaning that the speed of light [sic!] is independent
		of the polarization of the wave fields.
		The Maxwell--Lorentz equations for a point charge satisfy 
		(P),(W),(M),(D), but not (E).}
in ``one of those amusing cases of serendipity in theoretical physics'' 
(\cite{BiBiONE}, p.37). 

                \subsection{The Maxwell--Born--Infeld field equations}

 The Maxwell--Born--Infeld field equations, here for simplicity written only for a 
single (negative) point charge, consist of Maxwell's general field equations with 
the point charge source terms $\rho(t,\sV)\equiv -e \delta_{\QV(t)}(\sV)$ and 
$\jV(t,\sV)\equiv - e\delta_{\QV(t)}(\sV)\dot{\QV}(t)$, 
\begin{alignat}{1}
\textstyle
        \frac{1}{c}\pddt{\BMBIpt(t,\sV)}
&= \label{eq:MBIdotB}
        -\nab\times\EMBIpt(t,\sV) \, ,
\\
\textstyle
         \frac{1}{c}\pddt{\DMBIpt(t,\sV)}
&= \label{eq:MBIdotD}
        +\nab\times\HMBIpt(t,\sV)  
	+ 4\pi \textstyle{\frac{1}{c}} e\delta_{\QV(t)}(\sV)\dot{\QV}(t)\,,
\\
        \nab\cdot \BMBIpt(t,\sV)  
&= \label{eq:MBIdivB}
        0\, ,
\\
        \nab\cdot\DMBIpt(t,\sV)  
&=\label{eq:MBIdivD}
       - 4 \pi e \delta_{\QV(t)}(\sV) \, ,
\end{alignat}
together with the electromagnetic ``aether law''\footnote{Note that Born and Infeld viewed 
	their nonlinear relationship between $\EV$, $\HV$ on one side and 
	$\BV$, $\DV$ on the other no longer as a constitutive law in the sense 
	of Maxwell, but as the \emph{electromagnetic law of the classical vacuum}.}
of Born and Infeld \cite{BornInfeldBb},
\begin{alignat}{1}
&{\EMBIpt} 
= \label{eq:FOLIeqEofBD}
\frac{{\DMBIpt}-\frac{1}{b^2}{\BMBIpt}\times({\BMBIpt}\times{\DMBIpt})}
 {\sqrt{ 
1+\frac{1}{b^2}(|{\BMBIpt}|^2+|{\DMBIpt}|^2)+\frac{1}{b^4}|{\BMBIpt}\times{\DMBIpt}|^2}
  }\,,
\\
&{\HMBIpt} 
 = \label{eq:FOLIeqHofBD}
\frac{{\BMBIpt}-\frac{1}{b^2}{\DMBIpt}\times({\DMBIpt}\times {\BMBIpt})}
 {\sqrt{ 
1+\frac{1}{b^2}(|{\BMBIpt}|^2+|{\DMBIpt}|^2)+\frac{1}{b^4}|{\BMBIpt}\times{\DMBIpt}|^2}
  }\,,
\end{alignat}
where $b\in (0,\infty)$ is \emph{Born's field strength}, a hypothetical new 
``constant of nature,'' which Born determined \cite{BornB, BornInfeldA, BornInfeldBa, BornInfeldBb} as follows.

	\subsubsection{Born's determination of $b$}

	In the absence of any charges, these \emph{source-free} Maxwell--Born--Infeld field 
equations conserve the field energy, the field momentum, and the field angular 
momentum, given by the following integrals, respectively (cf. \cite{BiBiONE}),
\begin{equation}
\hskip-5pt
  \Ef 
= \label{eq:FIELDenergyMBI}
  \frac{b^2}{4\pi}\int_{\mathbb R^3}\!
\Bigl(\!{\sqrt{1 + \textstyle{\frac{1}{b^2}}(|{\BMBI}|^2 + |{\DMBI}|^2) 
   + \textstyle{\frac{1}{b^4}}|{\BMBI}\times {\DMBI}|^2}} -1\Bigr)(t,\sV) \drm^3s,
\end{equation}
\begin{equation}
  \PVf 
= \label{eq:FIELDimpulsMBI}
  \frac{1}{4\pi c} \int_{\mathbb R^3} (\DMBI\times\BMBI)(t,\sV) \, \drm^3s\, , 
\end{equation}
\begin{equation}
  \LVf
= \label{eq:FIELDdrehimpulsMBI}
  \frac{1}{4\pi c} \int_{\mathbb R^3} \sV\times(\DMBI\times\BMBI)(t,\sV) \,\drm^3s\, .
\end{equation}
	Supposing that these integrals retain their meanings also in the presence 
of sources, Born computed the energy of the field pair $(\BV_{\mathrm{Born}},\DV_{\mathrm{Born}})$, 
with $\BV_{\mathrm{Born}}= \vect{0}$ and
\begin{equation}
{\DV}_{\mathrm{Born}}\bigl(\sV\bigr) 
=
{\DV}_{\mathrm{Coulomb}}\bigl(\sV\bigr) 
\equiv
- e {\sV\;}/{|\sV|^3},
\end{equation}
which is the unique electrostatic finite-energy solution to the 
Maxwell--Born--Infeld equations with a single\footnote{The only other 
		explicitly known electrostatic solution with point charges
		was found by Hoppe \cite{HoppeA}, 
		describing an infinite crystal with finite energy per charge \cite{GibbonsA}.} 
negative point charge source at the origin of (otherwise) empty space; 
see \cite{PryceB}, \cite{ecker}.
	The field energy of Born's solution is
\begin{equation}
\Ef\bigl({\vect{0}},{\DV}_{\mathrm{Born}}\bigr) 
=
\frac{b^2}{4\pi}
\int_{\Rset^3} \Big(
\sqrt{1 +\textstyle{\frac{1}{b^2}}|\DV_{\mathrm{Born}}(\sV)|^2} -1\Big)\drm^3s
= 
\textstyle{\frac{1}{6} }\Beta\!\left({\textstyle{\frac{1}{4},\frac{1}{4}}}\right) 
\textstyle{\sqrt{be^3}},
\end{equation}
where $\Beta(p,q)$ is Euler's Beta function; numerical evaluation gives
\begin{equation}
\textstyle{\frac{1}{6} }\Beta\!\left({\textstyle{\frac{1}{4},\frac{1}{4}}}\right) 
\approx 
1.2361.
\end{equation}
	Finally, inspired by the idea of the later 19th century that the electron's 
inertia a.k.a. mass has a purely electromagnetic origin, Born \cite{BornB}
now argued that $\Ef\bigl({\vect{0}},{\DV}_{\mathrm{Born}}\bigr) = \me c^2$,
thus finding for his field strength constant
\begin{equation}
 b_{_{\mathrm{Born}}}
= \label{eq:BORNconstantCOMPUTED} 
 36 \Beta\!\left({\textstyle{\frac{1}{4},\frac{1}{4}}}\right)^{-2} m^2 c^4 e^{-3} \,.
\end{equation}

	Subsequently Born and Schr\"odinger \cite{BornErwin}  argued that this value has to
be revised because of the electron's magnetic moment, and they came up with a very 
rough, purely magnetic estimate.
	Born then asked Madhava Rao \cite{RaoRING} to improve on their estimate by computing
the energy of the electromagnetic field of a charged, stationary circular current 
density,\footnote{This corresponds to a ring singularity in both the electric and 
		magnetic fields.}
but Rao's computation is too approximate to be definitive.

	We will have to come back to the determination of $b$ at a later time.

	\subsubsection{The  status of (P), (W), (M), (E), (D)}

	The Maxwell--Born--Infeld field equations \emph{formally} satisfy the postulates 
(P), (W), (M), (E), (D), and there is no other set of field equations which does so.
	However, in order to qualify as a proper mathematical realization of 
(P), (W), (M), (E), (D) they need to \emph{generically} have well-behaved solutions.
	In this subsubsection I briefly review what is rigorously known about generic solutions.
  
\smallskip
\noindent
\hskip1.4truecm
\textbf{Source-free fields}

\noindent
	For the special case of source-free fields the mentioning of point 
charge sources in (E) is immaterial.
	The finite-energy requirement remains in effect, of course.
 
	Brenier \cite{Brenier} has recently given a very ingenious proof that the source-free 
electromagnetic field equations \refeq{eq:MBIdotB}--\refeq{eq:FOLIeqHofBD} are hyperbolic and 
pose a Cauchy problem; see also \cite{SerreC}.
	The generic existence and uniqueness of global classical solutions realizing 
(P), (W), (M), (E), (D) for initial data which are sufficiently \emph{small} 
in a Sobolev norm was only recently shown,\footnote{This result had been 
		claimed in \cite{ChaeHuh}, but their proof contained a fatal error.}
by Speck \cite{SpeckMBI}, who also extended his result to the general-relativistic 
setting \cite{SpeckEMBI}.

	The restriction to small initial data is presumably vital because the works by 
Serre \cite{SerreA} and Brenier \cite{Brenier} have shown that arbitrarily regular 
plane wave initial data can lead to formation of a singularity in finite time. 
	Now, plane wave initial data trivially have an infinite energy, but in his Ph.D. 
thesis Speck also showed that the relevant plane-wave initial data can be suitably 
cut off to yield finite-energy initial data which still lead to a singularity in
finite time.
	The singularity is  a divergent field-energy density, not of the shock-type.
	However, it is still not known whether the initial data that lead to a singularity 
in finite time form an open neighborhood. 
	If they do, then formation of a singularity in finite time is a generic phenomenon 
of source-free Maxwell--Born--Infeld field evolutions.
	In this case it is important to find out how large, in terms of $b$, the field 
strengths of the initial data are allowed to be in order to launch a unique global evolution.
	Paired with empirical data this information should yield valuable bounds on $b$.

\smallskip
\noindent
\hskip1.4truecm
\textbf{Fields with point charge sources}

\noindent
	Alas, hardly anything is rigorously known about the Maxwell--Born--Infeld field 
equations with one (or more) point charge source term(s) for \emph{generic} smooth 
subluminal motions $t\mapsto \QV(t)$.
	Hopefully in the not too distant future it will be shown that this Cauchy problem 
is locally well-posed, thereby realizing (P), (W), (M), (E), (D) at least for short times. 
	A global well-posedness result would seem unlikely, given the 
cited works of Serre and Brenier.

	Only for the special case where all point charges remain at rest there are generic
existence and uniqueness results for \emph{electrostatic} solutions.
	By applying a Lorentz boost these electrostatic 
solutions map into unique  \emph{traveling} electromagnetic solutions;\footnote{Incidentally,
		{traveling} electromagnetic solutions satisfying (P), (W), (M), (E), (D) 
		cannot be source-free if they travel at speeds less than $c$ \cite{KiePLA2011}.}
of course, this says nothing about solutions for generic subluminal motions.

	The first generic electrostatic results were obtained for charges with sufficiently small magnitudes
in unbounded domains with boundary; see \cite{KlyMik}, \cite{Kly}.
	Recently it has been proved \cite{KieMBIinCMPprep} that a unique {electrostatic} 
solution realizing (P), (W), (M), (E), (D) exists for arbitrary placements, signs and 
magnitudes of arbitrarily (though finitely) many point charges in $\Rset^3$; 
the solutions have $C^\infty$ regularity away from the point charges.\footnote{In \cite{GibbonsA}
		it is suggested that such a result would follow from the results on maximal
		hypersurfaces described in \cite{BartnikMINIconf}. 
	Results by Bartnik and Simon \cite{BartnikSimon} and by Bartnik \cite{BartnikB} 
		are indeed important ingredients to arrive at our existence and regularity result, 
		but in themselves not sufficient to do so.}

	Incidentally, the existence of such electrostatic solutions can be perplexing.
	Here is Gibbons (p.19 in \cite{GibbonsA}): 
``[W]hy don't the particles accelerate under the influence of the mutual forces?
	The reason is that they are pinned to their fixed position ... by external forces.''  
	A more sober assessment of this situation will be offered in our next section.

    \section{Classical theory of motion}
	We now turn to the quest for a well-defined classical law of motion for the point charge
sources of the electromagnetic Maxwell--Born--Infeld fields.
	In the remainder of this presentation I assume that the Cauchy problem for the
Maxwell--Born--Infeld field equations with point charge sources which move smoothly 
at subluminal speeds is \emph{generically} locally well-posed.

	\subsection{Orthodox approaches: a critique}
	In the beginning there was an intriguing conjecture,
that because of their nonlinearity the Maxwell--Born--Infeld field equations \emph{alone} 
would yield a (locally) well-posed Cauchy problem for \emph{both}, the fields \emph{and} 
the point charges.
	In a sense this idea goes back to the works by Born and Infeld \cite{BornB, BornInfeldBb}
who had argued that the law of motion is already contained in the differential law of field 
energy-momentum conservation obtained as a consequence of the source-free equations (or, equivalently,
as a local consequence of the Maxwell--Born--Infeld field equations away from point sources).
         A related sentiment can also be found in the work of Einstein, Infeld, and Hoffmann 
\cite{EinsteinInfeldHoffmann}, who seemed to also derive further inspiration from Helmholtz' 
extraction of the equations of point vortex motions out of Euler's fluid-dynamical equations. 

	However, if this conjecture were correct, then the existence and uniqueness of well-behaved
electrostatic field solutions with fixed point charges, announced in the previous section, would
allow us to immediately dismiss the Maxwell--Born--Infeld field equations as unphysical. 
	For example, consider just two identical point charges \emph{initially at rest} and
far apart, and consider field \emph{initial data} identical to the pertinent \emph{electrostatic}
two-charge solution.
	If the field equations alone would uniquely determine the (local) future evolution of 
both, fields and point charges, they inevitably would have to produce the unique
electrostatic solution with the point charges remaining at rest, whereas 
a physically acceptable theory must yield that these two charges begin to 
(approximately) perform a degenerate Kepler motion, moving away from each other along a straight line.

	The upshot is: either the Maxwell--Born--Infeld field equations alone do form a
complete classical theory of electromagnetism, and then this theory is \emph{unphysical},
or they are incomplete as a theory of electromagnetism, in which case they may very well 
be part of a \emph{physically acceptable} classical theory of electromagnetism.
	As I've stated at the end of the previous section, I expect that it will be shown 
that the Cauchy problem for the Maxwell--Born--Infeld field equations is locally well-posed
for \emph{generic} prescribed smooth subluminal motions of their point charge sources.
	Assuming this to pan out, the Maxwell--Born--Infeld field equations would have to 
be complemented by an additional set of dynamical equations which 
characterize the classical physical motions amongst all possible ones.

	While we have been using electrostatic solutions in three dimensions to argue for the 
incompleteness of the Maxwell--Born--Infeld field equations as a classical theory of electromagnetism, 
the first person, apparently, to have realized the flaw in the ``intriguing conjecture''
was Born's son-in-law Maurice Pryce who, after finding analogous electrostatic solutions in 
two dimensions \cite{PryceA}, wrote (\cite{PryceD}, p.597): ``It was clear from the 
start of the New Field Theory (although not fully appreciated in I and II 
[i.e. \cite{BornB} and \cite{BornInfeldBb}])
that the motion of the charges was not governed by the field equations alone, and that some further 
condition had to be added.''
	But then Pryce continued: ``It was also clear from physical considerations of conservation of 
energy and momentum what this condition had to be; namely, that the total force (...) on each charge 
must vanish.''
        His proposal is truthful to the revival, by Born and Infeld \cite{BornInfeldA}, 
\cite{BornInfeldBb}, of the old idea that the inertial mass of the electron
has a purely electromagnetic origin, and therefore it follows Abraham's and Lorentz' proposal 
for the equation of motion in a purely electromagnetic model (``the total force vanishes'').
	However, since, as mentioned earlier, in \cite{AppKieAOP} it was shown that the purely electromagnetic 
Abraham-Lorentz model is overdetermined and generically has no solution, we have the benefit of hindsight
and should be apprehensive of Pryce's proposal.
	Yet, until \cite{AppKieAOP} nobody had found anything mathematically
suspicious in Abraham's and Lorentz' manipulations up to second order,\footnote{Of course, 
		the third order term always was a source of much discussion and confusion.
	And as candidates for a fundamental model of the physical electron these classical approaches 
	were abandoned long ago.}
which got repeated verbatim at least until \cite{JacksonBOOKb}, so it should come as no surprise that Pryce's
adaptation of the Abraham-Lorentz reasoning to the Born-Infeld setting was accepted by most peers. 
	In particular, Schr\"odinger \cite{ErwinDUBLINb} picked up on Pryce's proposal and tried to 
push the approximate evaluation by including more terms in an expansion using spherical harmonics.

	Interestingly, Dirac \cite{DiracBI} used a refinement of Born's original  approach 
\cite{BornB} (recanted in \cite{BornInfeldBa, BornInfeldBb}) to arrive at the Newtonian law of motion 
\refeq{eq:NLeOFm}, \refeq{eq:ELPpv}, with $\EMBIpt$ and $\BMBIpt$ for, respectively, $\EV$ and $\BV$ 
in \refeq{eq:NLeOFm}, then went on to define the Lorentz force with the total fields through 
regularization which involved manipulations of the energy-momentum-stress tensor approach used 
also by Pryce. 
	Dirac also had a ``non-electromagnetic mass $M$'' for $\me$ in \refeq{eq:ELPpv}, but 
remarked: ``We may have $M=0$, which is probably the case for an electron.''

	I will now first explain why the Lorentz self-force still cannot be 
well-defined, and why postulating field energy-momentum conservation is not going to help, neither
in the basic version as proposed first by Born and Infeld \cite{BornInfeldBb} nor in the amended
manner discussed by Pryce \cite{PryceD}.
	Our analysis will also bear on some more recent discussions, e.g. \cite{GibbonsA}.
	
	I will then describe a well-defined Hamilton--Jacobi law of motion by following 
\cite{KieJSPaMBI}, in fact improving over the one proposed in  \cite{KieJSPaMBI}.

	\subsection{The inadequacy of the Lorentz self-force concept}
  
	\subsubsection{Ambiguity of the regularized Lorentz self-force}
	Even if the Maxwell--Born--Infeld field equations with point charge sources 
have solutions realizing (P), (W), (M), (E), (D) for generic smooth subluminal motions, 
as conjectured to be the case, the formal expression for the Lorentz force, 
``$\EV(t,\QV(t)) + {\scriptstyle{\frac{1}{c}}}\dot\QV(t) \times\BV(t,\QV(t))$,'' 
is still undefined \emph{a priori} when $\EV(t,\sV)=\EMBIpt(t,\sV)$ and 
$\BV(t,\sV)=\BMBIpt(t,\sV)$ are the total electric and magnetic fields.
	Also, it cannot be defined at the locations of the point charge(s) by a limit
$\sV\to\QV(t)$  of $\EMBIpt(t,\sV)$ and $\BMBIpt(t,\sV)$ at time $t$, 
because these fields cannot be continuously extended to $\sV=\QV(t)$.
	Again one can hope to define the total Lorentz force by taking a point limit 
$\varrho(\sV|\QV(t))\to\delta_{\QV(t)}(\sV)$ of a Lorentz force field 
$\EMBIpt(t,\sV) + {\scriptstyle{\frac{1}{c}}}\dot\QV(t) \times\BMBIpt(t,\sV)$ 
averaged over some normalized regularizing density $\varrho(\sV|\QV(t))$. 
	While in contrast to our experience with this procedure when applied to the solutions of the 
Maxwell--Lorentz field equations one may now obtain a finite result, because the point charge's 
self-energy is now finite, this is still not a satisfactory definition because even such a finite 
result will still depend on the specific details of the regularization $\varrho$ and its removal. 
	For instance, Dirac \cite{DiracBI} proposed that the procedure should be such as to yield a 
\emph{vanishing} contribution from the discontinuous part of the electromagnetic fields, but this
to me seems as arbitrary as proposing that the regularization should produce the minimal or the maximal
possible force, or some other ``distinguished'' vector.
	Since Dirac invoked the energy-momentum-stress tensor for his regularization argument, I will 
comment on a few more details in the next subsection.
	Here the upshot is: unless a \emph{compelling principle} is found which resolves the 
ambiguity of the self-field contribution to the total Lorentz force on a point charge, the Lorentz 
force has to be purged from the list of fundamental classical concepts.

	\subsubsection{The postulate of local energy-momentum conservation} 

	As I've claimed already, there exists a unique \emph{electrostatic} solution of the 
Maxwell--Born--Infeld field equations with $N$ point charges \emph{remaining at rest} at their initial positions.
	Since the integrals of field energy and field momentum (and field angular momentum) are all conserved 
by any static solution of the Maxwell--Born--Infeld field equations, this shows that simply postulating their 
conservation does not suffice to produce the right motions.
	Of course, postulating only the conservation of these global quantities 
is an infinitely much weaker requirement than postulating \emph{detailed local balance}
		$\partial_\nu \TMBIpt^{\mu\nu} = 0^\mu$ \emph{everywhere},
where $\TMBIpt^{\mu\nu}$ are the components of the symmetric energy(-density)-momentum(-density)-stress 
tensor of the {electromagnetic} Maxwell--Born--Infeld field; the local law for field angular-momentum 
conservation follows from this one and needs not to be postulated separately. 
	Note that in the absence of point charges the law $\partial_\nu \TMBIpt^{\mu\nu} = 0^\mu$
is a consequence of the field equations and not an independent postulate.
	Only in the presence of point charges, when $\partial_\nu \TMBIpt^{\mu\nu} = 0^\mu$ is
valid a priori only \emph{away from} these point charges, its continuous extension into the 
locations of the point charges amounts to a new postulate, indeed.
	Yet also this ``local conservation of field energy-momentum'' postulate does not deliver 
what it promises, and here~is~why.

	Consider the example of fields with two identical point charges (charged $-e$, say).
	The charges move smoothly at subluminal speed, and suppose the fields solve the 
Maxwell--Born--Infeld field equations.
	Introducing the electromagnetic stress tensor of the Maxwell--Born--Infeld fields, 
\begin{alignat}{1}
\textstyle
\hskip-23pt
         \ThetaMBIpt
= \label{eq:ThetaMBI} 
\textstyle\frac{1}{4\pi}\Big[
\EV\otimes\DV + \HV\otimes\BV  
- b^2 \Bigl(\!{\sqrt{1 + \textstyle{\frac{1}{b^2}}(|{\BV}|^2 + |{\DV}|^2) 
		   + \textstyle{\frac{1}{b^4}}|{\BV}\times {\DV}|^2}} -1\Bigr)\II \Big],
\end{alignat}
with $\EV=\EMBIpt$ etc., when viewing $\TMBIpt^{\mu\nu}(t,\sV)$ as a $t$-family of distributions 
over $\Rset^3$, 
the space components of $\partial_\nu \TMBIpt^{\mu\nu}$ then yield the \emph{formal} identity 
\begin{alignat}{1}
\textstyle
\hskip-23pt
\frac{1}{4\pi c}
 &\textstyle\pddt{(\DMBIpt\times\BMBIpt)}(t,\sV) - \nab\cdot \ThetaMBIpt(t,\sV)
= \label{eq:dnuTmunuDELTA} \\
&\textstyle \qquad\qquad
e [\mbox{``$\EMBIpt(t,\QV_1(t))$''} + 
	\frac1c \dot\QV_1(t)\times\mbox{``$\BMBIpt(t,\QV_1(t))$''}] \delta_{\QV_1(t)}(\sV)  \ +
\notag\\
&\textstyle\qquad\qquad
 e [\mbox{``$\EMBIpt(t,\QV_2(t))$''} + 
	\frac1c \dot\QV_2(t)\times\mbox{``$\BMBIpt(t,\QV_2(t))$''}] \delta_{\QV_2(t)}(\sV) ,
\notag
\end{alignat}
where the quotes around the field symbols in the second line of \refeq{eq:dnuTmunuDELTA}
remind us that the electric and magnetic fields are generally ill-defined at $\QV_1$ 
and $\QV_2$, indicating that our problem has caught up with us again.
	At this point, to restore complete electromagnetic energy-momentum conservation,
Pryce \cite{PryceD} and Schr\"odinger \cite{ErwinDUBLINb} rationalized that in effect
one has to postulate r.h.s.\refeq{eq:dnuTmunuDELTA}$=\nullV$. 
	This in essence is what Pryce meant by ``the total force on each charge must vanish.''
	(The law of energy conservation is dealt with analogously.)
	They go on and extract from this postulate the familiar Newtonian test particle 
law of motion in weak applied fields, the rest mass of a point charge given by its 
electrostatic field energy$/c^2$.

	For Dirac, on the other hand, allowing an extra non-electromagnetic mass $M$, 
r.h.s.\refeq{eq:dnuTmunuDELTA}$\not=\nullV$ because  a time-dependent extra kinetic energy associated
with $M$ had to be taken into account, and only total energy-momentum should be conserved.
        Hence he made a different postulate to remove the ambiguity highlighted by the quotes
around the electromagnetic field at the locations of the point charges.
	Eventually also Dirac obtained the familiar Newtonian test particle law of motion, but 
when $M=0$ at the end of the day, the rest mass of the point charge became its purely electrostatic 
field energy$/c^2$, too. 

	To exhibit the \emph{subtle mathematical and conceptual issues} 
in the reasonings of Pryce, Schr\"odinger, and Dirac, we return (briefly) to 
the special case where the two charges are initially at rest,
with  electrostatic field initial data.

	In this particular case, by continuous and \emph{consistent} extension into $\QV_1$ and $\QV_2$, 
we find that $\pddt{(\DMBIpt\times\BMBIpt)}(t,\sV)=\nullV$ for all $t$ and $\sV$.
	Now pretending r.h.s.\refeq{eq:dnuTmunuDELTA} were well-defined as a 
vector-valued distribution, i.e. if ``$\EMBIpt(0,\QV_1)$'' and ``$\EMBIpt(0,\QV_2)$'' were actual 
vectors (note that $\dot\QV_k=\vect{0}$ now), we can use Gauss' divergence theorem to actually compute 
these vectors.
	Thus, integrating \refeq{eq:dnuTmunuDELTA} over \emph{any} smooth, 
bounded, simple open domain $\Lambda$ containing, say, $\QV_1$ but not $\QV_2$ then yields
$-e$``$\EMBIpt(0,\QV_1)$''$=\int_{\partial\Lambda}\ThetaMBIpt\cdot\nV \drm\sigma$.
	The surface integral at the right-hand side is well-defined for any such $\Lambda$, but since
the left-hand side of this equation is independent of $\Lambda$, also the surface integral must
be independent of $\Lambda$.
	Happily it \emph{is} independent of $\Lambda$ (as long as $\Lambda$ does not contain $\QV_2$) because for the
electrostatic field the distribution $\nab\cdot \ThetaMBIpt(0,\sV)$ is supported only at $\QV_1$ and $\QV_2$.
	In the absence of any explicit formula for the electrostatic two-point solution one so-far relies on 
an approximate evaluation. 
Gibbons \cite{GibbonsA} suggests that for large separations between the point charges
the answer is Coulomb's force formula.
	(In \cite{PryceA} the two-dimensional electrostatic field  with two point charges
is computed exactly, and the closed line integral of the stresses around a charge shown to yield Coulomb's
formula for distant charges.)
	If proven rigorously correct in three dimensions, and presumably it can be shown rigorously,
this would seem to invalidate Pryce's (and 
Schr\"odinger's) line of reasoning that ``$\EMBIpt(0,\QV_1)$'' and ``$\EMBIpt(0,\QV_2)$'' could be 
postulated to vanish. 

	However, Pryce and Schr\"odinger were no fools. 
	They would have pointed out that what we just explained in the previous paragraph would be an
unambiguous definition of ``$\EMBIpt(0,\QV_1)$'' and ``$\EMBIpt(0,\QV_2)$'' for the 
\emph{electrostatic field solution} to the Maxwell--Born--Infeld field equations, because we used that 
for all $t$ and $\sV$  we have $\pddt{(\DMBIpt\times\BMBIpt)}(t,\sV)=\nullV$ for the electrostatic solution
(actually, this is only needed for $t\in(-\epsilon,\epsilon)$).
	But, they would have continued, the electrostatic solution is unphysical and has to be ruled out,
and this is precisely one of the things which postulating ``$\EMBIpt(0,\QV_1)$''$=\nullV$ and 
``$\EMBIpt(0,\QV_2)$''$=\nullV$ for the \emph{physical solution} accomplishes, thanks to 
the mathematical results of the previous paragraph which show that for the \emph{physical solution} to the 
Maxwell--Born--Infeld field equations with electrostatic initial data for fields and particles one cannot have 
$\pddt{(\DMBIpt\times\BMBIpt)}(0,\sV)=\nullV$ for all $\sV$.
	This in turn implies that the point charges for the \emph{physical solution} cannot remain at
rest but are accelerated by the electrostatic forces so defined.
	Furthermore, they would have insisted, since the physical 
$\frac{1}{4\pi c}\pddt{(\DMBIpt\times\BMBIpt)}(t,\sV)$ has to exactly 
offset the distribution $\nab\cdot \ThetaMBIpt(t,\sV)$,
one obtains an equation of motion for the positions of the point charges for all times.

	Brilliant! 
	But does it work?
	There are two issues to be addressed. 
	
	First, there is the issue as to the definition of the forces on the point charges in general 
dynamical situations. 
	Since the surface integrals $\int_{\partial\Lambda_k}\ThetaMBIpt(t,\sV)\cdot\nV \drm\sigma$, with $\Lambda_k$ 
containing only $\QV_k(t)$, will now generally depend on $\Lambda_k$, one can at best define the force on the
$k$th charge at time $t$ by taking the limit $\Lambda_k\to\{\QV_k(t)\}$, provided the limit exists and is
independent of the particular shapes of the shrinking $\Lambda_k$.
	Whether this is possible is a mathematical issue, regarding the behavior of the field solutions near
the point charges that move at generic, smooth subluminal speeds.
	Thus, $\DMBIpt(t,\sV)$ and $\HMBIpt(t,\sV)$ must not diverge stronger than 
$1/|\sV-\QV_k(t)|^2$ at each $\QV_k(t)$; to get nontrivial forces, they must in fact 
diverge exactly at this rate in leading order.
	This is an open problem, but it is not unreasonable to assume, as Pryce did, that this will be proven true,
at least for sufficiently short times.

	The second issue is more problematic.
	Namely, the distribution $\nab\cdot \ThetaMBIpt(t,\sV)$ would, for
sufficiently short times $t>0$ after the initial instant, be of the form 
$\fV(t,\sV) + \sum_k \fVN_k(t)\delta_{\QV_k(t)}(\sV)$, where $\fV(t,\sV)$ is a regular
force density field, while $\fVN_k(t)$ is the above defined force vector on the $k$th point charge. 
	The field $\fV(t,\sV)$ will be precisely offset by the regular part of 
$\frac{1}{4\pi c}\pddt{(\DMBIpt\times\BMBIpt)}(t,\sV)$ thanks to the local conservation law of electromagnetic
energy-momentum away from the charges, as implied by the field equations alone.
	Thus, in order to get an equation of motion in line with Newtonian classical physical notions,
the singular (distributional) part of $\pddt{(\DMBIpt\times\BMBIpt)}(t,\sV)$ at $\QV_k(t)$ now 
must be of the form $\delta_{\QV_k(t)}(\sV)$ times a vector which depends on $\QV_k(t)$ and its
first two time-derivatives --- note that also the initial source terms for the field equations
require $\QV_k(0)$ and $\dot\QV_k(0)$ to be prescribed, suggesting a \emph{second-order} equation
of motion for the $\QV_k(t)$.
	In particular, for the initial data obtained from the electrostatic field solution, with 
charges initially at rest and very far apart, the ``physical'' 
$\frac{1}{4\pi c}\pddt{(\DMBIpt\times\BMBIpt)}(0,\sV)$ has to be asymptotic to
$\sum_k \me\ddot\QV_k(0)\delta_{\QV_k(0)}(\sV)$ as $|\QV_1(0)-\QV_2(0)|\to\infty$, 
if it is to reproduce the physically correct equation of slow and gently accelerated Kepler
motions of two physical electrons in the classical regime.

	I do not see how the second issue could be resolved favorably. 
	In fact, it should be worthwhile to try to come up with a proof that the putative 
equation of motion is overdetermined, in the spirit of \cite{AppKieAOP}, but I haven't
tried this yet. 

	But how could Born, Infeld, Pryce, and Schr\"odinger, all have convinced themselves
that this procedure will work? 
	It is illuminating to see the answer to this question, because it will sound an alarm.

	Consider once again the electrostatic initial data with two identical point charges initially 
at rest and far apart.
	Let $(\Rset^2\equiveg)\Sigma\subset\Rset^3$ be the symmetry plane for this electrostatic 
field, and let $\Lambda_k$ be the open half-space containing $\QV_k(0)$, $k=1,2$; thus, 
$\Rset^3=\Lambda_1\cup\Lambda_2\cup \Sigma$.
	Then, by integrating their postulated equation of motion over $\Lambda_k$, and with $k+k'=3$, 
we find
\begin{alignat}{1}
\hskip-30pt
{\textstyle\frac{1}{4\pi c}\Ddt}\!\int_{\Lambda_k}\!\!\!{(\DMBIpt\times\BMBIpt)}(t,\sV)\drm^3s\Big|_{t=0}
=
\int_{\partial\Lambda_k}\!\!\!\ThetaMBIpt(0,\sV)\cdot\nV \drm\sigma
\sim 
e^2
\textstyle\frac{\QV_k(0)-\QV_{k'}(0)\;}{|\QV_k(0)-\QV_{k'}(0)|^3}
\end{alignat}
with ``$\sim$'' as the distance between the charges tends $\to\infty$ (the asymptotic result is 
assumed to be true, and presumably rigorously provable as I've written already).
	Pryce and his peers next argued that for gently accelerated motions we should be allowed
to replace the field momentum integral $\frac{1}{4\pi c}\int_{\Lambda_k}\!{(\DMBIpt\times\BMBIpt)}(t,\sV)\drm^3s$ 
by $\mf(\QV_1,\QV_{2}) \gamma_k(t)\dot\QV_k(t)$, where $\gamma_k^2(t) = 1/(1-\textstyle|\dot\QV_k(t)|^2/c^2)$
and 
$\mf(\QV_1,\QV_{2})c^2 =  \frac{b^2}{4\pi} \int_{\Lambda_k} 
\bigl(\!{\scriptstyle\sqrt{\textstyle 1+b^{-2}|\DMBIpt|^2}}-1\bigr)(0,\sV) \drm^3s$,
with $\QV_k$ standing for $\QV_k(0)$.
	Lastly, we should have $\DMBIpt(0,\sV-\QV_k)\to {\DV}_{\mathrm{Born}}\bigl(\sV\bigr)$
as $|\QV_k-\QV_{k'}|\to\infty$, with (the relevant) $\QV_k$ at the origin,
and in this sense, and with $b=b_{\mathrm{Born}}$, we would then also
have $\mf(\QV_1,\QV_{2})\to \me $ as $|\QV_1-\QV_2|\to\infty$.
	Thus, in this asymptotic regime, at the initial time, the reasoning of Pryce and his peers yields
\begin{alignat}{1}
{\textstyle\Ddt}\big(\me\gamma_k(t)\dot\QV_k(t)\big)\big|_{t=0}
=
e^2
\textstyle\frac{\QV_k(0)-\QV_{k'}(0)\;}{|\QV_k(0)-\QV_{k'}(0)|^3}.
\end{alignat}

	Surely this looks very compelling, but it is clear that the  heuristic replacement of 
$\frac{1}{4\pi c}\int_{\Lambda_k}\!{(\DMBIpt\times\BMBIpt)}(t,\sV)\drm^3s$ by 
$\mf(\QV_1,\QV_{2}) \gamma_k(t)\dot\QV_k(t)$ as just explained is only a first approximation.
	By going one step further Schr\"odinger \cite{ErwinDUBLINb} argued that the (in)famous 
third-order radiation reaction term will appear, and so, if the approximation were consistent, 
we would now need a third initial condition, namely $\ddot\QV_k(0)$.
	We are ``back to square one'' with our problems. 
	One might argue that the third-order term is not yet the consistent approximation and
invoke the Landau--Lifshitz reasoning to get an effective second-order equation.
	However, going on to higher orders would successively bring in higher and higher derivatives
of $\QV_k(t)$.
	This looks just like the situation in the old purely electromagnetic classical electron 
theory of Abraham and Lorentz.
	All alarm bells should be going off by now, because their 
equation of motion is overdetermined \cite{AppKieAOP}.

	Dirac, on the other hand, obtains as putatively exact equation of motion
\begin{alignat}{1}
{\textstyle\Ddt}\big(M\gamma_k(t)\dot\QV_k(t)\big)
= \label{eq:DIRACeqOFmot}
e_k
\Big[\EMBIptreg(t,\QV_k(t)) + {\scriptstyle{\frac{1}{c}}}\dot\QV_k(t) \times\BMBIptreg(t,\QV_k(t))\Big],
\end{alignat}
where the superscripts $^{\rm{reg}}$ indicate his regularization procedure, which
also involves the integration of the stress tensor over a small domain $\Lambda_k$ 
containing $\QV_k(t)$, plus Gauss' theorem, followed by the limit $\Lambda_k\to\{\QV_k(t)\}$; 
however, Dirac uses this only to split off a singular term from the ill-defined Lorentz force, arguing 
that the remaining electromagnetic force field is regular; this field enters in \refeq{eq:DIRACeqOFmot}.
	As far as I can tell, Dirac's remainder field is generally still singular; the critical
passage is on the bottom of page 36 in \cite{DiracBI}.

	I end here with a comment on Dirac's suggestion that $M=0$ for an electron.
	In this case, even with a regular force field, his \refeq{eq:DIRACeqOFmot} 
would be overdetermined, because setting the coefficient of the highest derivative 
equal to zero amounts to a singular limit.

    \subsubsection{On Newton's law for the rate of change of momentum}

	I have argued that no matter how you cut the cake, the Lorentz force 
formula r.h.s.\refeq{eq:NLeOFm} for the electromagnetic force on a point charge 
cannot be well-defined when $\EV$ and $\BV$ are the total fields.
	Since only the total fields can possibly be fundamental, while ``external field'' 
and ``self-field'' are only auxiliary notions, it follows that the Lorentz force cannot 
play a fundamental dynamical role.
	This now inevitably raises the question as to the status of Newton's law for the 
rate of change of momentum, $\dot\PV(t) = \fVN(t)$, of which \refeq{eq:NLeOFm} pretends
to be the particular realization in the context of classical electrodynamics.

	If one insists that Newton's law $\dot\PV(t) = \fVN(t)$ remains fundamental throughout
classical physics, including relativistic point charge motion coupled to the 
electromagnetic fields, then one is obliged to continue the quest for a well-defined 
expression for the fundamental electromagnetic force $\fVN$ on a point charge.

	The alternative is to relegate Newton's law $\dot\PV(t) = \fVN(t)$ to the
status of an effective law, emerging in the regime of relativistic charged test 
particle motions.
	In this case one needs to look elsewhere for the fundamental relativistic law 
of motion of point charges.
	Of course, the effective concept of the \emph{external Lorentz force} acting on a 
\emph{test} particle remains a beacon which any fundamental theory must keep in sight.

	This is the point of view taken in \cite{KieJSPaMBI}, and also here.

    \subsection{Hamilton--Jacobi theory of motion}
        \label{sec:HamJac}

	Although we have not only abandoned \refeq{eq:NLeOFm} but actually Newton's 
$\dot\PV(t) = \fVN(t)$ altogether, for now we will hold on to the second one of 
the two equations \refeq{eq:NLeOFm}, \refeq{eq:ELPpv} of the \emph{formal} 
relativistic Newtonian law of motion. 
	But then, since \refeq{eq:ELPpv} expresses the rate of change of position, 
$\dot\QV(t)$, in terms of the ``mechanical momentum'' vector $\PV(t)$, we need 
to find a new type of law which gives us $\PV(t)$ in a well-defined manner.
	Keeping in mind the moral that ``formal manipulations are not to be trusted until 
they can be vindicated rigorously,'' in this section we will argue that 
Hamilton--Jacobi theory supplies a classical law for $\PV(t)$ which in fact 
is well-defined, provided the solutions realizing (P), (W), (M), (E), (D) 
of the Maxwell--Born--Infeld field equations with point charge sources are 
as well-behaved for generic smooth subluminal motions as they are for Born's static 
solution, at least locally in time.
 
	Indeed, the electric field ${\EV}_{\mathrm{Born}}$ associated to 
$({\BV}_{\mathrm{Born}},{\DV}_{\mathrm{Born}})$, Born's static field pair for a 
single (negative) point charge at the origin, is undefined at the origin but 
uniformly bounded elsewhere; it exhibits a \emph{point defect} at $\sV=\vect{0}$.
	For $\sV\neq \vect{0}$, it is given by
${\EV}_{\mathrm{Born}}\bigl(\sV\bigr) = -\nab{\phi}_{\mathrm{Born}}({\sV})$, where
\begin{equation}
\phi_{\mathrm{Born}}({\sV}) 
=  
- \sqrt{be}
 \int_{|{\scriptscriptstyle\sV}|\sqrt{{b}/{e}}}^\infty\; \frac{\drm{x}}{\sqrt{1+ x^4}} 
\,.
\label{eq:BornsElectricPot}
\end{equation}
	Note, $\phi_{\mathrm{Born}}({\sV}) \sim -e |\sV|^{-1}$ for $|\sV|\gg\sqrt{e/b}$, and 
$\lim_{|\sV|\downarrow{0}}\phi_{\mathrm{Born}}(\sV) 
 = :\phi_{\mathrm{Born}}(\vect{0})<\infty$.
 So, away from the origin the electrostatic potential $\phi_{\mathrm{Born}}(\sV)$ is 
(even infinitely) differentiable, and it can be Lipschitz-continuously extended into 
the origin. 
	We conjecture that this regularity is \emph{typical} for the electromagnetic 
\emph{potentials} of the Maxwell--Born--Infeld fields for \emph{generic} smooth 
subluminal point source motions, in the sense that it should be so in the 
Lorenz--Lorentz gauge (see below), and remain true under any subsequent \emph{smooth}
gauge transformation. 
	Gauge transformations with less regularity would have to be ruled out.

	\subsubsection{The electromagnetic potentials}

	Given a solution $t\mapsto (\BMBIpt,\DMBIpt)(t,\sV)$ of the Maxwell--Born--Infeld 
field equations for some smooth subluminal motion $t\mapsto\QV(t)$ of a point charge 
(or several of them), we can algebraically compute the field pair 
$t\mapsto (\BMBIpt,\EMBIpt)(t,\sV)$ from the Born--Infeld ``aether law.''
	For any such map $t\mapsto (\BMBIpt,\EMBIpt)(t,\sV)$, we define the magnetic vector potential 
$\AMBIpt(t,\sV)$ and the electric potential $\phiMBIpt(t,\sV)$ in terms of the following PDE.
	Namely, $\AMBIpt(t,\sV)$ satisfies the evolution equation 
\begin{equation}
\textstyle
        \frac{1}{c}\pddt{\AMBIpt(t,\sV)}
= \label{eq:FOLIeAmagn}
        - {\nab}\phiMBIpt(t,\sV) - {\EMBIpt(t,\sV)}
\end{equation}
and the constraint equation
\begin{equation}
        {\nab}\times  {\AMBIpt}(t,\sV)
= \label{eq:FOLIcAmagn}
          {\BMBIpt}(t,\sV)\, ,
\end{equation}
while the evolution of $\phiMBIpt(t,\sV)$ is governed by
\begin{alignat}{1}
\textstyle
 \pddt \phiMBIpt(t,\sV) 
= \label{eq:LLgauge}
- {\nab} \cdot  {\AMBIpt}(t,\sV),
\end{alignat}
unconstrained by any other equation.

	Equation \refeq{eq:LLgauge} is known as the  V.Lorenz--H.A.Lorentz 
gauge condition (see \cite{HawkingEllisBOOK, JacksonOkun}
postulated here for no other reason than that it is simple, invariant under the 
Poincar\'e group, and \emph{presumably} compatible with our regularity conjecture
for the potentials.
	While it renders the Maxwell--Lorentz field equations with prescribed 
point sources as a decoupled set of non-homogeneous wave equations for the
four-vector field $(\phiML,{\AML})$, readily solved by the Li\'enard--Wiechert 
potentials \cite{lienard, Wiechert}, \refeq{eq:LLgauge} achieves no such 
simplification for the Maxwell--Born--Infeld equations with point sources.

	The Lorenz--Lorentz condition fixes the gauge freedom of the relativistic 
four-vector potential field $\big(\phi,\AV\big)(t,\sV)$ to some extent, yet the 
equations \refeq{eq:FOLIeAmagn}, \refeq{eq:FOLIcAmagn}, and \refeq{eq:LLgauge} 
are still invariant under the gauge transformations
\begin{alignat}{1}
  \phi(t,{\sV}) 
& 
\to \phi(t,{\sV}) - {\textstyle\pddct} \Upsilon(t,{\sV}),
\\
    {\AV} (t,{\sV})
& 
\to  {\AV}(t,{\sV}) + \nab \Upsilon (t,{\sV}),
\label{eq:gaugetrPHIandA}
\end{alignat}
with any relativistic scalar field $\Upsilon: {\Rset}^{1,3}\to{\Rset}$ 
satisfying the wave equation
\begin{equation}
{\textstyle\pddctSQUARE} \Upsilon  (t,{\sV}) = {\nab^2}  \Upsilon  (t,{\sV}),
\label{eq:waveEQupsilon}
\end{equation}
with $\nab^2 = \Delta$, the Laplacian on $\Rset^3$.
	Since a (sufficiently regular) solution of  \refeq{eq:waveEQupsilon} 
in $\Rset_+\times \Rset^3$ is uniquely determined by the initial data for $\Upsilon$ and its 
time derivative $\pddt\Upsilon$, the gauge freedom that is left concerns the initial 
conditions of $\phi,{\AV}$. 

	\subsubsection{Canonical momenta of point defects with intrinsic mass $m$}

	As per our (plausible but as of yet unproven) hypothesis, for generic smooth 
subluminal point charge motions the electromagnetic potential fields $\phiMBIpt$ 
and $\AMBIpt$ have Lipschitz continuous extensions to all of space, at any time $t$.
	In the following we shall always mean these extensions when we speak of the 
electromagnetic-field potentials.
	With our electromagnetic-field potentials unambiguously defined at each location of a 
field point defect, we are now able to define the so-called canonical momentum of a 
point defect of the electromagnetic fields associated with a point charge moving along 
a smooth trajectory $t\mapsto\QV(t)$ with subluminal speed.
	Namely, given a smooth trajectory $t\mapsto\QV(t)$, consider \refeq{eq:ELPpv} 
though now with ``intrinsic inert mass $m$'' in place of $\me$.
Inverting  \refeq{eq:ELPpv} with intrinsic inert mass $m$ we obtain the ``intrinsic'' 
momentum of the point defect,
\begin{equation}
\PV(t) 
= \label{eq:MOMENTUMasFCTofVELOCITY}
m \frac{\dot{\QV}(t)}{\sqrt{1 - {|\dot{\QV}(t)|^2}/{c^2}}}.
\end{equation}
	Also, per our conjecture, 
${\AMBIpt}(t,{\QV}(t))$ is well-defined at each $t$, and so, for a negative charge,
the \emph{canonical momentum}
\begin{equation}
\PiV(t) 
:= \label{eq:canonicalMOMENTUMnegCHARGE}
{\PV}(t)  -   {\textstyle{\frac{1}{c}}} e {\AMBIpt}(t,{\QV}(t))
\end{equation}
is well-defined for all $t$.
 \newpage

	We now turn \refeq{eq:canonicalMOMENTUMnegCHARGE} around and, for a negative point 
charge, take
\begin{equation}
\PV(t) 
= \label{eq:pVisPVpluseA}
{\PiV}(t)  +   {\textstyle{\frac{1}{c}}} e {\AMBIpt}(t,{\QV}(t))
\end{equation}
as the formula for $\PV(t)$ that has to be coupled with \refeq{eq:ELPpv}.
	Thus, next we need to find an expression for $\PiV(t)$. 
	Precisely this is supplied by Hamilton--Jacobi theory.


	\subsubsection{Hamilton--Jacobi laws of motion}

 In Hamilton--Jacobi theory of single-point motion one introduces the 
\emph{single-point configuration space} of \emph{generic positions} 
$\qV\in\Rset^3$ of the point defect. 
 A \emph{Hamilton--Jacobi law of motion} consists of two parts: (i) an
\emph{ordinary differential equation} for the actual position $\QV(t)$
of the point defect, equating its actual velocity with the evaluation --- 
at its actual position --- of a velocity field on configuration space; 
(ii) a \emph{partial differential equation} for this velocity field.
 The correct law should reduce to the test particle theory in certain regimes, 
so we begin with the latter.

\medskip
\noindent
\hskip1.4truecm
\textbf{The test charge approximation: all well again, so far!}

\smallskip
\noindent
       In advanced textbooks on mathematical classical physics \cite{ThirringBOOKa} one 
finds the equations of relativistic Hamilton--Jacobi theory for test charge motion in 
the potentials $\AV(t,\sV)={\AMM}(t,\sV)$ and $\phi(t,\sV)=\phiMM(t,\sV)$  for the 
 \emph{actual field} solutions $\EMM(t,\sV)$ and $\BMM(t,\sV)$ of the Maxwell--Maxwell 
field equations, serving as ``external'' fields.
       This reproduces the test particle motions 
computed from \refeq{eq:NLeOFm}, \refeq{eq:ELPpv} with $\EV=\EMM$ and $\BV=\BMM$.
 This setup is locally well-defined, for ``externally generated potentials'' are independent 
of where and how the test charge moves, and so they can be assumed to be smooth 
functions of space and time.
 Provided the solutions $\EMBI(t,\sV)$ and $\BMBI(t,\sV)$ of the \emph{source-free} 
Maxwell--Born--Infeld field equations are smooth, the Hamilton--Jacobi law of test 
particle motion remains locally well-defined if we set
$\AV(t,\qV)={\AMBI}(t,\qV)$ and $\phi(t,\qV)=\phiMBI(t,\qV)$, the 
generic-$\qV$-evaluation of these source-free Maxwell--Born--Infeld potentials.

\smallskip
\noindent
\centerline{\textit{The relativistic Hamilton--Jacobi guiding equation}}

\vskip1pt
\noindent
Suppose the actual canonical momentum of the point test charge is given by
\begin{equation}
\PiV(t)
= \label{eq:PisgradS}
{\nablaq} \SHJ(t,{\QV}(t)),
\end{equation}
where $\qV\mapsto \SHJ(t,\qV)$ is a time-dependent differentiable scalar field on configuration space.
         By virtue of \refeq{eq:PisgradS}, \refeq{eq:MOMENTUMasFCTofVELOCITY}, \refeq{eq:pVisPVpluseA}, 
with $\AMBIpt(t,\QV(t))$ replaced by ${\AMBI}(t,\QV(t))$, we can eliminate $\PiV(t)$ in favor of 
$\nablaq \SHJ(t,\QV(t))$ which, for a negative test charge of mass $m$, yields the 
\emph{relativistic Hamilton--Jacobi guiding equation} 
\begin{equation}
  \frac{1}{c}
\frac{\drm{\QV(t)}}{\drm{t}}
=\label{eq:STANDARDhamjacLAWofMOTIONtest}
\frac{{\nablaq} \SHJ(t,{\QV(t)}) +   {\textstyle{\frac{1}{c}}} e{\AMBI}(t,{\QV(t)})}
{\sqrt{m^2c^2+
   {|{\nablaq}\SHJ(t,{\QV(t)})+{\textstyle{\frac{1}{c}}}e{\AMBI}(t,{\QV(t)})|}{}^2}}
\,.
\end{equation}

\smallskip
\noindent
\centerline{\textit{The relativistic Hamilton--Jacobi partial differential equation}}
 
\smallskip
\noindent
 The requirement that the test charge velocity is the space component of a 
(future-directed) four-velocity vector divided by the relativistic $\gamma$ factor quite
naturally leads to the following \emph{relativistic Hamilton--Jacobi partial differential equation},
\begin{equation}
\textstyle
  \frac{1}{c} \pddt \SHJ(t,\qV) 
= \label{eq:HamJacPDEtestCHARGE}
-{\sqrt{m^2c^2+{|\nablaq\SHJ(t,\qV)+{\textstyle{\frac{1}{c}}}e {\AMBI}(t,\qV)|}^2}} + 
  {\textstyle{\frac{1}{c}}} e \phiMBI(t,\qV) .
\end{equation}

\medskip
\noindent
\centerline{\textit{Lorentz and Weyl invariance}}

\smallskip
\noindent
 As to Lorentz invariance, any solution to \refeq{eq:HamJacPDEtestCHARGE} obviously satisfies
\begin{equation}
\big({\textstyle{\frac{1}{c}\pddt}}\SHJ(t,\qV) 
-{\textstyle{\frac{1}{c}}}e\phiMBI(t,\qV)\big)^2 -
{\big|\nablaq \SHJ(t,\qV) + {\textstyle{\frac{1}{c}}}e {\AMBI}(t,\qV)\big|}^2
 = \label{eq:HamJacPDEsquaredTEST} 
m^2c^2,
\end{equation}
a manifestly relativistically Lorentz scalar equation. 

 Although $\SHJ(t,\qV)$ is a scalar configuration spacetime field, it cannot be 
gauge-invariant, for the four-vector field $\big(\phi,\AV\big)(t,\sV)$ is not; 
recall that the Lorenz--Lorentz gauge condition alone does not fix the potentials 
completely.
 Instead, if the potentials $(\phi,\AV)$ are transformed under the  
gauge transformations \refeq{eq:gaugetrPHIandA} with any relativistic 
scalar field $\Upsilon: {\Rset}^{1,3}\to{\Rset}$ satisfying the wave equation 
\refeq{eq:waveEQupsilon}, then, for a negative charge, $\SHJ$ needs to be transformed as
\begin{equation}
  \SHJ(t,{\qV})
\to \label{eq:gaugetrS}
\SHJ(t,{\qV})  -   {\textstyle{\frac{1}{c}}} e \Upsilon(t,{\qV}).
\end{equation}
   This gauge transformation law also holds more generally for $\Upsilon$ not satisfying 
\refeq{eq:waveEQupsilon}, meaning a change of gauge from Lorenz--Lorentz to something else.

\medskip
\noindent
\centerline{\textit{Many-body test charge theory}}
 
\smallskip
\noindent
 The generalization to many point charges with either sign is obvious.
 Since test charges do not ``talk back'' to the ``external'' potentials, there is 
a guiding equation \refeq{eq:STANDARDhamjacLAWofMOTIONtest} coupled with a partial 
differential equation \refeq{eq:HamJacPDEtestCHARGE} for the guiding velocity field 
\emph{for each test charge}.
 Of course, they are just identical copies of the single-particle equations, yet it is 
important to keep in mind that the many-body theory is to be formulated on many-body 
configuration space.

\medskip
\noindent
\hskip1.2truecm
\textbf{Upgrading test particle motions: self-force problems d\'ej\`a vu!}

\smallskip
\noindent
	Since the electromagnetic potentials for the \emph{actual} electromagnetic 
Maxwell--Born--Infeld fields with point charge sources are supposedly defined 
everywhere, it could now seem that in order to get a well-defined theory of motion 
of their point charge sources all that needs to be done is to replace $\AMBI(t,\qV)$ 
and $\phiMBI(t,\qV)$ by $\AMBIpt(t,\qV)$ and $\phiMBIpt(t,\qV)$ in 
\refeq{eq:HamJacPDEtestCHARGE}, which yields the partial differential equation
\begin{equation}
\textstyle
  \frac{1}{c} \pddt \SHJ(t,\qV) 
= \label{eq:HamJacPDEtestCHARGEupgrade}
-{\sqrt{m^2c^2 +{|\nablaq \SHJ(t,\qV)+{\textstyle{\frac{1}{c}}}e{\AMBIpt}(t,\qV)|}^2}}
 +   {\textstyle{\frac{1}{c}}} e \phiMBIpt(t,\qV) ,
\end{equation}
and to replace $\AMBI(t,\QV(t))$ by $\AMBIpt(t,\QV(t))$ in 
\refeq{eq:STANDARDhamjacLAWofMOTIONtest} to get the guiding equation
\begin{equation}
  \frac{1}{c}
\frac{\drm{\QV(t)}}{\drm{t}}
=\label{eq:STANDARDhamjacLAWofMOTIONtestupgrade}
\frac{{\nablaq} \SHJ(t,{\QV(t)}) +   {\textstyle{\frac{1}{c}}} e{\AMBIpt}(t,{\QV(t)})}
  {\sqrt{m^2c^2 +{|{\nablaq} \SHJ(t,{\QV(t)})+{\textstyle{\frac{1}{c}}} e{\AMBIpt}(t,{\QV(t)})|}{}^2}}
\,.
\end{equation}
 So the \emph{actual} electromagnetic potentials as functions of space and time are 
evaluated at the generic position $\qV$ in \refeq{eq:HamJacPDEtestCHARGEupgrade} and 
at the actual position $\QV(t)$ in  \refeq{eq:STANDARDhamjacLAWofMOTIONtestupgrade}.

 Note that almost all flow lines of the gradient field  $\nablaq \SHJ(t,\qV)$ would 
still correspond to test particle motions in the actual $\phiMBIpt(t,\sV)$ and 
$\AMBIpt(t,\sV)$ potential fields (simply because almost all generic positions 
$\qV$ are not identical to the actual position $\QV(t)$ of the point charge source 
of the actual $\phiMBIpt(t,\sV)$ and $\AMBIpt(t,\sV)$), so
one may hope that by suitably iterating the given actual motion one can make precisely 
one of these test particle motions coincide with the actual motion --- which is meant 
by ``upgrading test-particle motion.'' 

 However, this does not lead to a well-defined theory of motion of point charge 
sources!
 The reason is that $\phiMBIpt(t,\sV)$ and $\AMBIpt(t,\sV)$ have non-differentiable 
``kinks'' at $\sV=\QV(t)$.
  The function $\SHJ(t,\qV)$ picks up this non-differentiability at $\qV=\QV(t)$ 
through \refeq{eq:HamJacPDEtestCHARGEupgrade}.
  More precisely, \refeq{eq:HamJacPDEtestCHARGEupgrade} is only well-defined 
\emph{away} from the actual positions of the point charges.
 Trying to extend the definition of $\nablaq \SHJ(t,\qV)$ to the actual positions 
now leads pretty much to the same mathematical problems as encountered when trying 
to define the ``Lorentz self-force'' on the point charge sources of the 
Maxwell--Born--Infeld field equations.
 In particular, we could regularize the actual potentials $\phiMBIpt(t,\sV)$ and 
$\AMBIpt(t,\sV)$ by averaging, thereby obtaining a regularized ``upgraded 
test-particle Hamilton--Jacobi theory'' which does yield the actual ``regularized 
motion'' amongst all ``regularized test particle motions'' as a nonlinear fixed 
point problem. 
 Unfortunately, subsequent removal of the regularization generally does not yield 
a unique limit, so that any so-defined limiting theory of point charge motion would,
once again, not be well-defined.

 Fortunately, Hamilton--Jacobi theory offers another option. 
 Recall that for the non-relativistic problem of motion of $N$ widely separated 
point charges interacting through their Coulomb pair interactions, 
Hamilton--Jacobi theory yields a gradient flow on $N$-particle configuration 
space of which \emph{each flow line represents a putative actual trajectory} 
of the $N$ body problem: there are no test particle trajectories!
 In this vein, we should focus on a formulation of Hamilton--Jacobi theory which 
``{parallel-processes}'' putative actual point charge motions.

\medskip
\noindent
\hskip1.4truecm
\textbf{Parallel processing of putative actual motions: success!}

\smallskip
\noindent
 While nontrivial motions in a strictly non-relativistic Coulomb problem without ``external'' 
fields can occur only when $N\geq 2$ (Kepler motions if $N=2$), a system 
with a single point charge source for the electromagnetic Maxwell--Born--Infeld fields 
generally should feature non-trivial motions on single-particle configuration space
because of the dynamical degrees of freedom of the electromagnetic fields.
 So in the following we focus on the $N=1$ point charge problem, although 
eventually we have to address the general $N$-body problem.

 Setting up a Hamilton--Jacobi law which ``parallel-processes'' putative actual
single point source motions in the Maxwell--Born--Infeld field equations 
is only possible if there exists a generic velocity field on configuration space 
(here: for a negative point charge), denoted by $\vV(t,\qV)$, which varies smoothly with $\qV$
and $t$, and which is related to the family of putative actual motions by
the guiding law
\begin{equation}
\frac{\drm{\QV(t)}}{\drm{t}}
=\label{eq:guidingLAWofMOTIONpp}
\vV(t,\QV(t))
\,,
\end{equation}
yielding the actual position $\QV(t)$ for each actual initial position $\QV(0)$.
 
	Assuming such a velocity field exists, one next needs to construct configuration space fields
${\phi}_{1}(t,{\qV})$ and ${\AV}_{1}(t,{\qV})$ which are ``generic-$\qV$-sourced''
potential fields $\phi^\sharp(t,\sV,\qV)$ and $\AV^\sharp(t,\sV,\qV)$ evaluated at 
$\sV=\qV$, their generic point
source;\footnote{The notation is inherited from the $N$-point-charge problem. 
        In that case there are fields $\phi^\sharp(t,\sV,\qV_1,...,\qV_N)$ etc.  which, when 
		evaluated at $\sV=\qV_k$, give configuration space fields $\phi_k(t,\qV_1,...,\qV_N)$ 
		etc.
	For a system with a single point charge, $k=1$.}
i.e.
\begin{alignat}{1}
{\phi}_{1}(t,{\qV})\equiv \phi^\sharp(t,\qV,\qV) 
\qquad \mbox{and}\qquad 
{\AV}_{1}(t,{\qV}) \equiv \AV^\sharp(t,\qV,\qV),
\end{alignat}
see \cite{KieJSPaMBI}.
 The ``canonical'' set of partial differential equations for 
${\phi}^\sharp(t,\sV,\qV)$, ${\AV}^\sharp(t,\sV,\qV)$, and their 
derived fields, which are compatible with the Maxwell--Born--Infeld field equations 
for the actual fields of a single negative point charge, reads
\begin{alignat}{1}
\hskip-30pt
{\textstyle\pddct}\phi^\sharp(t,\sV,\qV)
&=  \label{eq:FOLIeqPHIoneNEG}
- {\textstyle\frac{1}{c}}\vV(t,\qV)\!\cdot\! \nablaq \phi^\sharp(t,\sV,\qV)
- \nab\!\cdot\! \AV^\sharp(t,\sV,\qV),
\\
\hskip-30pt
\textstyle
\pddct {\AV}^\sharp(t,\sV,\qV)
&=\label{eq:FOLIeqAoneNEG}
- {\textstyle\frac{1}{c}}
\vV(t,\qV)\!\cdot\!\nablaq  {\AV}^\sharp(t,\sV,\qV)
- \nab \phi^\sharp(t,\sV,\qV) -  {\EV}^\sharp(t,\sV,\qV),
\\
\hskip-30pt
\textstyle
\pddct {\DV}^\sharp(t,\sV,\qV)
&= \label{eq:FOLIeqDoneNEG}
- {\textstyle\frac{1}{c}}\vV(t,\qV)\!\cdot\!\nablaq
 {\DV}^\sharp(t,\sV,\qV)\!
+\! \nab\!\! \times\! {\HV}^\sharp(t,\sV,\qV)\!
+\! 4\pi e {\textstyle\frac{1}{c}}{\vV}(t,{\qV}) \delta_{{\qV}}({\sV});
\end{alignat}
furthermore, ${\DV}^\sharp(t,\sV,\qV)$ obeys the constraint equation\footnote{Since 
		the generic charge density $-e\delta_{\qV}(\sV)$ is $t$-independent
		and $\nablaq\delta_{{\qV}}({\sV})  = -\nab\delta_{{\qV}}({\sV})$,
		the reformulation of the continuity equation for charge conservation (in spacetime),
		$\pddt\rho^\sharp(t,\sV,\qV) =  
			- \vV(t,\qV)\cdot\nablaq \rho^\sharp(t,\sV,\qV) -\nab\cdot \jV^\sharp(t,\sV,\qV)$, 
		is an identity, not an independent equation.}
\begin{equation}
\nab\!\cdot\! {\DV}^\sharp(t,\sV,\qV)
=
 - 4\pi e \delta_{{\qV}}({\sV}).
\label{eq:FOLIeqDoneNEGconstraint}
\end{equation}
	The fields ${\EV}^\sharp(t,\sV,\qV)$ and ${\HV}^\sharp(t,\sV,\qV)$ in 
\refeq{eq:FOLIeqAoneNEG}, \refeq{eq:FOLIeqDoneNEG}
are given in terms of ${\DV}^\sharp(t,\sV,\qV)$ and ${\BV}^\sharp(t,\sV,\qV)$ 
in precisely the same way as the actual fields ${\EMBIpt}(t,\sV)$ and ${\HMBIpt}(t,\sV)$ 
are defined in terms of ${\DMBIpt}(t,\sV)$ and ${\BMBIpt}(t,\sV)$ 
through the Born--Infeld aether law \refeq{eq:FOLIeqEofBD}, \refeq{eq:FOLIeqHofBD}, while 
${\BV}^\sharp(t,\sV,\qV)$ in turn is given in terms of ${\AV}^\sharp(t,\sV,\qV)$ 
in the same way as the actual ${\BMBIpt}(t,\sV)$ 
is given in terms of the actual ${\AMBIpt}(t,\sV)$ in \refeq{eq:FOLIcAmagn}.
	It is straightforward to verify that by substituting the actual $\QV(t)$ 
for the generic $\qV$ in the ``generic-$\qV$-sourced'' $\sharp$-fields satisfying the above field equations, 
we obtain the actual electromagnetic potentials, fields, and charge-current densities satisfying 
the Maxwell--Born--Infeld field equations (in Lorenz--Lorentz gauge).
	That is, 
\begin{alignat}{1}
\phiMBIpt(t,\sV)\equiv \phi^\sharp(t,\sV,\QV(t)) 
\qquad \mbox{etc.}
\end{alignat}

	Next we need to stipulate a law for $\vV(t,\qV)$.

\medskip
\noindent
\centerline{\textit{The Hamilton--Jacobi velocity field}}

\smallskip
\noindent
	The na\"{\i}vely obvious thing to try is the generic velocity law 
\begin{equation}
\vV(t,\qV)
=\label{eq:STANDARDhamjacVfieldPP}
c \frac{         {\nablaq} \SHJ(t,{\qV}) +  {\textstyle{\frac{1}{c}}} e  {\AV}_{1}(t,{\qV})}
  {\sqrt{m^2c^2 +{|{\nablaq} \SHJ(t,{\qV})+ {\textstyle{\frac{1}{c}}} e  {\AV}_{1}(t,{\qV})|}{}^2}}
\,,
\end{equation}
corresponding to the Hamilton--Jacobi PDE
\begin{equation}
\textstyle
  \frac{1}{c} \pddt \SHJ(t,\qV) 
= \label{eq:HamJacPDEsourceCHARGEpp}
- {\sqrt{m^2c^2 +{|\nablaq \SHJ(t,\qV) + {\textstyle{\frac{1}{c}}}e {\AV}_1(t,\qV)|}^2}} + 
  {\textstyle{\frac{1}{c}}} e \phi_1(t,\qV) ,
\end{equation}
which replaces \refeq{eq:HamJacPDEtestCHARGE}.
	Since ${\AV}_1(t,{\QV(t)})= {\AMBIpt}(t,{\QV(t)})$, the guiding law
\refeq{eq:guidingLAWofMOTIONpp} with velocity field $\vV$ given by 
\refeq{eq:STANDARDhamjacVfieldPP} is superficially identical to 
\refeq{eq:STANDARDhamjacLAWofMOTIONtestupgrade}, yet note that $\SHJ$ in 
\refeq{eq:STANDARDhamjacLAWofMOTIONtestupgrade} is not the same $\SHJ$ as in 
\refeq{eq:STANDARDhamjacVfieldPP}, \refeq{eq:guidingLAWofMOTIONpp}
because ${\phi}(t,{\qV})$ and ${\AV}(t,{\qV})$ are now replaced by ${\phi}_{1}(t,{\qV})$ 
and ${\AV}_{1}(t,{\qV})$.
	Note also that our single-particle law of motion has a straightforward extension to
the $N$-body problem, which I also presented in \cite{KieJSPaMBI}.

	It is a reasonable conjecture that the maps $\qV\mapsto {\phi}_{1}(t,{\qV})$ and 
$\qV\mapsto{\AV}_{1}(t,{\qV})$ are \emph{generically} differentiable,\footnote{Normally, 
		a Cauchy problem is locally well-posed if there exists a unique solution, 
		locally in time, which depends Lipschitz-continuously on the initial data. 
	We here expect, and need, a little more regularity than what suffices for basic well-posedness.}
in which case one obtains the first well-defined self-consistent law of motion of a classical 
point charge source in the Maxwell--Born--Infeld field equations \cite{KieJSPaMBI}.
	It has an immediate generalization to $N$-particle systems. 

	It is straightforward to show that this law readily handles the simplest situation: the
trivial motion (i.e., rest) of the point charge source in Born's static solution.
	Note that no averaging or renormalization has to be invoked!

	Since the nonlinearities make it extremely difficult to evaluate the model in nontrivial 
situations, only asymptotic results are available so far.
	In \cite{KieJSPaMBI} and \cite{KieJSPbMBI} it is shown that a point charge 
in Maxwell--Born--Infeld fields which are ``co-sourced'' jointly by this charge and another 
one that, in a single-particle setup, is assumed to be immovable (a Born--Oppenheimer 
approximation to a dynamical two-particle setup), when the charges are far apart, 
carries out the Kepler motion in leading order, as it should.
	Moreover, at least formally one can also show that in general the slow motion and gentle
acceleration regime of a point charge is governed in leading order by a law of test charge 
motion as introduced at the beginning of this subsection.
	Whether this will pan out rigorously, and if so, whether the one-body setup yields physically 
correct motions if we go beyond the slow motion and gentle acceleration regime has yet to be 
established.

\subsubsection{Conservation laws: re-assessing the value of $b$}
\medskip

	In \cite{KieJSPaMBI} I explained that the system of Maxwell--Born--Infeld field equations
with a negative point charge source moving according to our parallel-processing Hamilton--Jacobi 
laws furnishes the following conserved total energy:
\begin{alignat}{1}
\hskip-5pt
  \cE
= \label{eq:FIELDenergyMBIpt}
\; &c{\sqrt{m^2c^2 +{|\nablaq \SHJ(t,\QV(t)) + {\textstyle{\frac{1}{c}}}e \AMBIpt(t,\QV(t))|}^2}}\quad + \\
\notag
&{\textstyle\frac{b^2}{4\pi}}\int_{\mathbb R^3}\!
\Bigl(\!{\sqrt{1 + \textstyle{\frac{1}{b^2}}(|{\BMBIpt}|^2 + |{\DMBIpt}|^2) 
   + \textstyle{\frac{1}{b^4}}|{\BMBIpt}\times {\DMBIpt}|^2}} -1\Bigr)(t,\sV) \drm^3s.
\end{alignat}
	In \cite{KieJSPaMBI} I had assumed from the outset that $m=\me$, but that was somewhat hidden
because of the dimensionless units I chose.
	The assumption $m=\me$ caught up with me when the \emph{total} rest mass of the 
point defect plus the electrostatic field around it, with $b=b_{\mathrm{Born}}$, became $2\me$. 
	With hindsight, I should have allowed the ``intrinsic mass of the defect'' $m$ to be a parameter, 
as I have done here, because then this bitter pill becomes bittersweet: there is a whole
range of combinations of $m$ and $b$ for which $\cE= \me c^2$; yet it is also evident that with $m>0$,
Born's proposal $b=b_{\mathrm{Born}}$ is untenable.
	More precisely, $b_{\mathrm{Born}}$ is an \emph{upper bound} on the admissible $b$ values
obtained from adapting Born's argument that the empirical rest mass of the physical electron should
now be the total energy over $c^2$ of a single point defect in its static field.

	What these considerations do not reveal is the relative distribution of mass between $m$ and $b$.
	My colleague Shadi Tahvildar-Zadeh has suggested that $m$ is possibly the only surviving
remnant of a general relativistic treatment, and thereby determined. 
	I come to general relativistic issues in the next subsection.

	Before I get to there, I should complete the listing of the traditional conservation laws.
	Namely, with a negative point charge, the total momentum,
\begin{alignat}{1}
\hskip-10pt
  \cP
= \label{eq:FIELDimpulsMBIpt}
\big[\nablaq \SHJ + {\textstyle{\frac{1}{c}}e \AMBIpt\big](t,\QV(t)) +
\frac{1}{4\pi c}} \int_{\mathbb R^3} (\DMBIpt\times\BMBIpt)(t,\sV) \, \drm^3s\, , 
\end{alignat}
and the total angular momentum, 
\begin{alignat}{1}
\hskip-10pt
  \cL
= \label{eq:FIELDdrehimpulsMBIpt}
\;\QV(t)\times \big[\nablaq \SHJ + {\textstyle{\frac{1}{c}}}e \AMBIpt\big](t,\QV(t)) +
{\textstyle \frac{1}{4\pi c}} \int_{\mathbb R^3} \sV\times(\DMBIpt\times\BMBIpt)(t,\sV) \,\drm^3s\,
\end{alignat}
are conserved as well.
	In addition there are a number of less familiar conservation laws, but this would lead us too
far from our main objective.

	\subsection{General-relativistic spacetimes with point defects}

	Ever since the formal papers by Einstein, Infeld, and Hoffmann \cite{EinsteinInfeldHoffmann},
there have been quite many attempts to 
prove that Einstein's field equations imply the equations of motion for ``point singularities.'' 
	Certainly they imply the evolution equations of continuum matter when the latter is the source of spacetime 
geometry, but as to true point singularities the jury is still out.
	For us this means a clear imperative to investigate this question rigorously when Einstein's field equations
are coupled with the Maxwell--Born--Infeld field equations of electromagnetism. 
	Namely, if Einstein's field equations imply the equations of motion for the point charges, as Einstein et al.
would have it, then all the developments described in the previous subsections have been in vain.
	If on the other hand it turns out that Einstein's field equations do not imply the equations 
of motion for the point charges, then we have the need for supplying such --- in that case the natural 
thing to do, for us, is to adapt the Hamilton--Jacobi type law of motion from flat to curved spacetimes.

	Fortunately, the question boils down to a static problem: Does the 
Einstein--Maxwell--Born--Infeld PDE system with two point charge sources have static, 
axisymmetric classically regular solutions away from the 
two worldlines of the point charges, no matter where they are placed? 
	If the answer is ``Yes,'' then Einstein's equations fail to deliver the equations of motion for the charges, for
empirically we know that two  \emph{physical} point charges in the classical regime would not remain motionless.
	Shadi Tahvildar-Zadeh and myself have begun to rigorously study this question.
	I hope to report its answer in the not too distant future. 

	Meanwhile, I list a few facts that by now are known and which make us quite optimistic.
	Namely, while the Einstein--Maxwell--Maxwell equations with point charges produce solutions 
with horrible naked singularities (think of the Reissner--Nordstr\"om spacetime with charge and mass 
parameter chosen to match the empirical electron data), the Einstein--Maxwell--Born--Infeld equations 
with point charge source are much better behaved.
	Tahvildar-Zadeh \cite{Shadi} recently showed that they not only admit a static spacetime corresponding to a 
single point charge whose ADM mass equals its electrostatic field energy$/c^2$, he also showed that the spacetime 
singularity is of the mildest possible form, namely a conical singularity. 
	Conical singularities are so mild that they lend us hope that the nuisance of ``struts'' between 
``particles,'' known from multiple-black-hole solutions of Einstein's equations, can be avoided. 
	Tahvildar-Zadeh's main theorem takes more than a page to state, after many pages of preparation.
	Here I will have to leave it at that.

	\section{Quantum theory of motion}
	Besides extending the classical flat spacetime theory to curved Lorentz manifolds, I have 
been working on its extension to the quantum regime. 
	In \cite{KieJSPbMBI} I used a method which I called \emph{least invasive quantization} of 
the one-charge Hamilton--Jacobi law for parallel processing of putative actual motions.
	Although I didn't see it this way at the time, by now I have realized that this least
invasive quantization can be justified elegantly in the spirit of the quest for unification
in physics! 
	\subsection{Quest for unification: least invasive quantization}
	If we accept as a reasonably well-established working hypothesis that dynamical physical theories 
derive from an action principle, we should look for an action principle for the Hamilton--Jacobi equation. 
	Because of the first order time derivative for $\SHJ$ such an action principle for the classical 
$\SHJ$ can be formulated only at the price of introducing a scalar companion field $\RHJ$
which complements $\SHJ$.
	To illustrate this explicitly it suffices to consider a representative, nonrelativistic 
Hamilton--Jacobi PDE, written as  $\pddt \SHJ(t,\qV) + H(\qV,\nab_{\qV} \SHJ(t,\qV)) = 0$. 
	Multiplying this equation by some positive function $R^2(t,\qV)$ and integrating over $\qV$ and $t$
(the latter over a finite interval $I$) gives the ``action'' integral

\vskip-17pt
\begin{equation}
A(R,\SHJ)
=\textstyle
\int_I\int_{\Rset^3}R^2(t,\qV)[\pddt \SHJ(t,\qV) + H(\qV,\nab_{\qV} \SHJ(t,\qV))]\drm^3\qV\drm t 
= 0.
\end{equation}

\vskip-7pt
\noindent
	Now replacing also $\SHJ$ by a generic $S$ in $A$ and seeking the stationary points of
$A(R,S)$, denoted by $\RHJ$ and $\SHJ$, under variations with fixed end points,
we obtain the Euler-Lagrange equations $\pddt \SHJ(t,\qV) + H(\qV,\nab \SHJ(t,\qV)) = 0$,
and $\pddt \RHJ^2(t,\qV) + \nab_{\qV}\cdot[\RHJ^2 \frac1m\nab_{\qV} \SHJ](t,\qV) = 0$. 
	Clearly, the $\SHJ$ equation is just the Hamilton--Jacobi equation we started from, while the 
$\RHJ$ equation is a passive evolution equation: a continuity equation.

	The passive evolution of $\RHJ$ somehow belies the fact that $\RHJ$ is needed to formulate 
the variational principle for $\SHJ$ in the first place.
	This suggests that $\RHJ$ is really a field of comparable physical significance to $\SHJ$.
	So in the spirit of unification, let's try to find a small modification of the  dynamics to
symmetrize the roles of $R$ and $S$ at the critical points.

	Interestingly enough, by adding an $R$-dependent penalty term (a 
Fisher entropy, $\propto\hbar^2$) to the action functional $A(R,S)$, one can obtain 
(even in the $N$-body case) a Schr\"odinger equation for its
critical points, denoted $\RQM e^{i\SQM/\hbar}=\psi$, where the  suffix $HJ$ has been replace by $QM$
to avoid confusion with ``$\RHJ e^{i\SHJ/\hbar}$.'' 
	The important point here is that the real and imaginary parts of $\RQM e^{i\SQM/\hbar}$ 
now satisfy a nicely symmetrical dynamics! 
	In this sense the $\RQM$ and $\SQM$ fields have been really \emph{unified} into a complex field $\psi$, 
whereas $\RHJ e^{i\SHJ/\hbar}$, while clearly complex, is not representing a unification of $\RHJ$ and $\SHJ$.
	Equally important: the guiding equation, and the ontology of points that move, is unaffected by 
this procedure!

\subsubsection{A de Broglie--Bohm--Klein--Gordon law of motion}

	The same type of argument works for the relativistic Hamilton--Jacobi theory and 
yields a Klein--Gordon equation. 
	The Klein--Gordon PDE for the complex scalar configuration space field $\psi(t,\qV)$ reads 
\begin{equation}
\big(i\hbar \pddct  +  e\textstyle\frac{1}{c} \phi_1\big)^2 \psi 
=  \label{eq:KleinGordonPDE}
m^2c^2 \psi + \big(\!-i\hbar\nablaq + e\textstyle\frac{1}{c} \AV_1\big)^2 \psi \,
\end{equation}
where $\phi_1$ and $\AV_1$ are the potential fields defined as in our parallel-processing 
single-charge Hamilton--Jacobi law.

	To wit, least invasive quantization does not affect the underlying purpose of the theory to 
provide a law of motion for the point defects. 
 For a Klein--Gordon PDE on configuration space the velocity field $\vV$ for the guiding equation 
$\dot\QV(t) = \vV(t,\QV(t))$ is now given by the ratio of quantum current vector density to density, 
$\jV^{\mathrm{qu}}(t,\qV)/\rho^{\mathrm{qu}}(t,\qV)$, with
\begin{equation}
\textstyle
\rho^{\mathrm{qu}}
=
\Im \left( \overline{\psi}\left(-\frac{\hbar}{mc^2} \pddt + i  \frac{e}{mc^2} \phi_1\right)\psi\right)
\,,
\qquad
\jV^{\mathrm{qu}}
=
 {\Im \left(\overline{\psi}\left( \frac{\hbar}{m}\nablaq + i  \frac{e}{mc} \AV_1\right)\psi\right)}
\label{eq:QrhojOFpsi}
\,,
\end{equation}
where $\Im$ means imaginary part, and $\overline{\psi}$ is the complex conjugate of $\psi$; thus
\begin{equation}
\vV(t,\qV)
\equiv
c \frac{\Im \left(\overline{\psi}\left(\hbar \nablaq + i e\textstyle\frac{1}{c} \AV_1\right)\psi\right)}
     {\Im \left(\overline{\psi}\left(-\hbar\pddct + i e\textstyle\frac{1}{c}\phi_1\right)\psi\right)}
(t,\qV)
\label{eq:QvelocityOFpsi}
\,.
\end{equation}
 This is a familiar de-Broglie--Bohm--Klein--Gordon law of motion, cf. \cite{duerrteufel,Holland}, except that
$\AV_1,\phi_1$ are not external fields, of course.

	\subsubsection{A de Broglie--Bohm--Dirac law of motion}

	It is only a small step from a Klein--Gordon to a Dirac equation for spinor-valued
$\psi$ coupled to the generic $q$-sourced potential fields for a negative charge,
\begin{equation}
i\hbar \pddct \psi 
=  \label{eq:DiracPDE}
mc \betaQ \psi
+
\alphaQ\cdot(-i\hbar\nablaq + e\textstyle\frac{1}{c} \AV_1) \psi 
- e\textstyle\frac{1}{c} \phi_1 \psi ;
\end{equation}
here $\alphaQ$ and $\betaQ$ are the familiar Dirac matrices. 
 The guiding equation for the actual point charge motion is still \refeq{eq:guidingLAWofMOTIONpp}, 
once again with $v = {j^{\mathrm{qu}}}/{\rho^{\mathrm{qu}}}$,
but now with the quantum density and quantum current vector density given by the Dirac expressions,
yielding the de Broglie--Bohm--Dirac guiding equation
\begin{equation}
  \frac{1}{c}
\frac{\drm{\QV(t)}}{\drm{t}}
=\label{eq:dBBDlawOFmotion}
 \frac{\psi^\dagger\alphaQ\psi}
     {\psi^\dagger\psi}(t,\QV(t))
\,,
\end{equation}
where $\Cset^4$ inner product is understood in the bilinear terms at the r.h.s.
 This is a familiar de-Broglie--Bohm--Dirac law of motion \cite{duerrteufel,Holland} except, once again, that
$\AV_1,\phi_1$ are not external fields.
	Presumably $\psi$ has to be restricted to an $\AV$-dependent ``positive energy subspace,'' which is
tricky, and we do not have space here to get into the details.

	\subsection{Born--Infeld effects on the Hydrogen spectrum} 

	The two-charge model with an electron and a nuclear charge in Born--Oppenheimer approximation is 
formally a dynamical one-charge model with an additional charge co-sourcing the Maxwell--Born--Infeld 
fields.
	It can be used to investigate Born--Infeld effects on the Hydrogen spectrum.

	The hard part is to find the electric potential $\phi^\sharp(\sV,\qV,\qV_n)$ of the electrostatic
Maxwell--Born--Infeld field of an electron at $\qV$ and the nucleus at $\qV_n=\vect{0}$ in otherwise
empty space.
	The conceptual benefits offered by the nonlinearity of the Maxwell--Born--Infeld 
field equations come at a high price: in contrast to the ease with which the general 
solution to the Maxwell--Lorentz field equations can be written down, there is no 
general formula to explicitly represent the solutions to the Maxwell--Born--Infeld 
field equations.
	So far only stationary solutions with regular sources can be written down 
systematically with the help of convergent perturbative series expansions \cite{CarKieCPDE, KieMBIinJMP}.

	In \cite{KieJSPbMBI} I presented an explicit integral formula for an approximation to  
$\phi^\sharp(\qV,\qV,\vect{0})=\phi_1(\qV)$.
	If the point charges are slightly smeared out and $b^{-2}$ is not too big, then
this formula gives indeed the electric potential for the leading order term in the perturbative
series expansion in powers of $b^{-2}$ for the displacement field $\DV$ developed in \cite{CarKieCPDE, KieMBIinJMP}.
	Assuming that the formula for the total electrostatic potential at the location of the electron
is giving the leading contribution also for point charges, Born--Infeld effects on the Schr\"odinger 
spectrum of Hydrogen were computed\footnote{The ground state energies as functions of Born's $b$ 
	parameter agree nicely in both numerical studies, but some of the excited states don't,
	hinting at a bug in our program.
	I thank Joel Franklin for pointing this out.}
in \cite{CarKiePRL,Franklin}.
	In \cite{Franklin} also the Dirac spectrum was studied.
	The interesting tentative conclusion from these studies is that Born's value of $b$ gives 
spectral distortions which are too large to be acceptable. 
	More refined two-body studies are still needed to confirm this finding, but the research clearly
indicates that atomic spectral data may well be precise enough to test the viability of the 
Born--Infeld law for electromagnetism.

	\section{Closing remarks}

	In the previous sections I have slowly built up a well-defined theory of motion for 
point defects in the Maxwell--Born--Infeld fields, both in the classical regime, 
using Hamilton--Jacobi theory, and also in the quantum regime, using wave equations 
without and with spin. 
	In either case the important notion is the parallel processing of motions, not test 
particle motions or their upgrade to a fixed point problem.

	Unfortunately, while the nonlinearity of the Maxwell--Born--Infeld equations makes 
the introduction of such laws of motion possible in the first place, it is  also an 
obstacle to any serious progress in computing the motions actually produced by these laws.
	But I am sure that it is only a matter of time until more powerful techniques are 
brought in which will clarify many of the burning open questions. 

	So far basically everything I discussed referred to the one-charge problem. 
	This is perfectly adequate for the purpose of studying the self-interaction problem 
of a point charge which lies at the heart of the problem of its motion. 
	But any acceptable solution to this self-interaction problem also has to be 
generalized to the $N$-charge situation, and this is another active field of inquiry.
	While the jury is still out on the correct format of the many charge theory, one 
aspect of it is presumably here to stay. 
	Namely, a many-charge formulation in configuration space clearly requires 
synchronization of the various charges; by default one would choose to work with a 
particular Lorentz frame, but any other choice should be allowed as well. 
	Actually, even the single-charge formulation I gave here tacitly uses the 
synchronization of the time components in the four-vectors $(ct,\sV)$ and $(q^0,\qV)$. 
	In the test charge approximation synchronization is inconsequential, but in this 
active charge formulation the many-charge law would seem to depend on the 
synchronization. 
	Whether the motion will depend on the foliation can naturally be investigated.
	Even if it does, the law of motion would not automatically be in conflict with 
Lorentz covariance.
	What is needed is simply a covariant foliation equation, as used in general 
relativity \cite{christodoulouklainermanBOOK}. 
	A distinguished foliation could be interpreted as restoring three-dimensionality to 
physical reality. 
	This would be against the traditional spirit of relativity theory, i.e. 
Einstein's interpretation of it as meaning that physical reality is 
four-dimensional, but that's OK. 
\newpage

\smallskip\noindent
\textbf{ACKNOWLEDGEMENT:} I am very grateful to the organizers
Felix Finster, Olaf M\"uller, Marc Nardmann, J\"urgen Tolksdorf, and Eberhard Zeidler
of the excellent ``Quantum Field Theory and Gravity'' conference in Regensburg (2010), 
for the invitation to present my research, and for their hospitality and support.
	The material reported here is based on my conference talk, but I have taken the
opportunity to address in more detail some questions raised in response to my presentation,
there and elsewhere.
	I also now mentioned the definitive status of some results that had been in the making
at the time of the conference.
	The research has been developing over many years, funded by  
NSF grants DMS-0406951 and DMS-0807705, which is gratefully acknowledged. 
	I benefited greatly from many discussions with my 
colleagues, postdocs and students, in particular: Walter Appel, Holly Carley, Sagun Chanillo,
Demetrios Christodoulou, Detlef D\"urr, Shelly Goldstein, Markus Kunze, Tim Maudlin, Jared Speck, 
Herbert Spohn, Shadi Tahvildar-Zadeh, Rodi Tumulka, Nino Zangh\`\i. 
	Lastly, I thank the anonymous referee for many helpful suggestions.


\end{document}